\documentclass[journal]{IEEEtran}

\usepackage{amssymb}
\usepackage{bbm}
\usepackage{mathbbol}
\usepackage{newtxmath}
\usepackage{amsmath}
\usepackage{graphicx}
\usepackage{subfigure}
\usepackage{caption}
\usepackage{algorithm}
\usepackage{booktabs}
\usepackage{algorithmic}
\usepackage[top=0.65in, bottom=0.65in, left=0.7in, right=0.7in]{geometry}
\usepackage{bm}
\usepackage{array}    
\usepackage{amsmath}  

\usepackage{subcaption} 
\usepackage{graphicx}   
\usepackage{graphicx}       
\usepackage{caption}        
\usepackage{subcaption}     

\begin{document}

\title{VQ-DSC-R: Robust Vector Quantized-Enabled Digital Semantic Communication With OFDM Transmission}

\author{Jianqiao Chen, Nan Ma, \textit{Member, IEEE}, Xiaodong Xu, \textit{Senior Member, IEEE},
    Tingting Zhu, Huishi Song, Chen Dong, Wenkai Liu, Rui Meng and Ping Zhang, \textit{Fellow, IEEE}
\thanks{
Jianqiao Chen, TingTing Zhu and HuiShi Song are with 
the ZGC Institute of Ubiquitous-X Innovation and Applications, and Beijing Key Laboratory of 6G DOICT converged and Cloud-Native Mobile Information Networks, Beijing 100876, China.
Nan Ma, Xiaodong Xu, Chen Dong, Wenkai Liu, Rui Meng and Ping Zhang are with the State Key Laboratory of Networking and Switching Technology, Beijing University of Posts and Telecommunications, Beijing 100876, China.  (Corresponding author: Nan Ma.)
(e-mail: \{chenjianqiao, zhutingting, songhuishi\}@zgc-xnet.com,  \{liuwenkai, manan, xuxiaodong, dongchen, liuwenkai, buptmengrui, pzhang\}@bupt.edu.cn)}
}

\maketitle
\thispagestyle{empty}
\begin{abstract}

Digital mapping of semantic features is essential for achieving interoperability between semantic communication and practical digital infrastructure. However, current research efforts predominantly concentrate on analog semantic communication with simplified channel models.
To bridge these gaps, we develop a robust vector quantized-enabled digital semantic communication (VQ-DSC-R) system built upon orthogonal frequency division multiplexing (OFDM) transmission. Our work encompasses the framework design of VQ-DSC-R, followed by a comprehensive optimization study.
Firstly, we design a Swin Transformer-based backbone for hierarchical semantic feature extraction, integrated with VQ modules that map the features into a shared semantic quantized codebook (SQC) for efficient index transmission. 
Secondly, we propose a differentiable vector quantization with adaptive noise-variance (ANDVQ) scheme to mitigate quantization errors in SQC, which dynamically adjusts the quantization process using K-nearest neighbor statistics, while exponential moving average mechanism stabilizes SQC training.
Thirdly, for robust index transmission over multipath fading channel and noise, we develop a conditional diffusion model (CDM) to refine channel state information, and design an attention-based module to dynamically adapt to channel noise.
The entire VQ-DSC-R system is optimized via a three-stage training strategy.
Extensive experiments demonstrate superiority of VQ-DSC-R over benchmark schemes, achieving high compression ratios and robust performance in practical scenarios.

\end{abstract}


\IEEEpeerreviewmaketitle

\section{Introduction}

\subsection{Background and Significance}

\IEEEPARstart{W}{ith} the large-scale interconnection of intelligent devices and the surge in wireless data traffic, the challenge of spectrum scarcity has become increasingly pressing, which poses significant obstacles for 6G wireless communication [1], [2]. To some extent, it is crucial to balance sustainability and performance for moving the 6G era [3]. However, existing communication technologies predominantly prioritize the precise transmission of each symbol, often overlooking the ultimate tasks and the intrinsic meaning conveyed by the data [4], [5]. The convergence of communication and AI (ComAI) framework is recently proposed to envision a co-evolving system where AI is embedded into the core communication processes, and communication infrastructures are inherently built to support intelligent functions [6]. In particular, it leverages semantic-level information to simplify and restructure signal processing, information cognition, and network organization. Therefore, semantic-level communication shifts the focus from precise bit delivery to conveying only the most relevant meaning, which provides a crucial advantage in extremely bandwidth-limited and power-constrained environments.

One of the enabling technologies for semantic communication is the deep joint source-channel coding (DJSCC), which employs artificial intelligence (AI) technologies to synergistically unify source coding and channel coding into a unified optimization framework [7]. 
Initial research efforts have studied that semantic features extracted by deep neural networks (DNNs) inherently adopt continuous magnitude representations (e.g., 32-bit floating-point values). In this way, it has revealed enhanced compression efficiency compared to conventional separated source channel coding (SSCC) approaches that combine source codecs with practical channel codecs [8]-[10]. However, transmitting such continuous-valued representations is either infeasible or highly inefficient in digital communication, which stems from the fundamental mismatch between continuous feature spaces and discrete digital signaling paradigms.

Unlike the analog semantic communication, digital semantic communication (DSC) that considers digital mapping of semantic features has been actively explored in recent studies. One of the widely-adopted approaches is scalar quantization (SQ), which independently maps semantic features to a finite set of discrete values for practical deployment [11]-[13]. 
However, the predefined SQ-based DSC is difficult to exploit correlations among different semantic features, which therefore hinders effective adaptation to the specific statistical distributions of the semantic features. To deal with it, a non-linear quantization module is proposed to efﬁciently quantize semantic features with trainable quantization levels [14]. In this case, the trainable quantization level requires additional parameter learning and gradient calculation, and the parameters need to be finely adjusted to avoid quantization error oscillation and confront non-convergence of training.

\subsection{Related Works and Contributions} 
Another research direction, centered on vector quantized (VQ)-enabled DSC, has received increasing interests [15]-[18]. 
In this way, it has the potential to adopt a trainable codebook for efficient representation of semantic features, which simultaneously enhances the interpretability of semantic communication.
The core commonality of these studies lies in jointly training a codebook with the semantic encoder and decoder for different modalities. 
However, they face fundamental issues of training codebook that need further consideration. 
Firstly, the continuous distribution of latent vectors is clustered into a finite codebook by VQ mechanism, which introduces the additional distortion termed quantization error. Secondly, the non-differentiable nature of widely-adopted nearest-neighbor assignment hinders gradient flow during backpropagation, causing the gradient collapse issue. Meanwhile, the codebook collapse, where a substantial fraction of the codebook's vectors converges to zero norm and remain unused, significantly degrades the representational ability of the codebook. 

Some studies have been conducted on these issues in the fields of computer science (CV) and natural language processing (NLP). 
We partition these efforts into two categories: (1) schemes that improve gradient backpropagation and (2) schemes that improve codebook capacity. For the former, a common solution to address the non-differentiability issue in quantization operations is to approximate gradients using the straight-through estimator (STE) [19]. 
However, it causes errors due to directly copying gradients from the decoder’s input and pasting them into the encoder’s output. 
Several recent studies have pursued solutions to the problem of gradient backpropagation of STE through the avoidance of deterministic quantization [20]-[22]. 
For the latter, the approaches to mitigate codebook collapse or under-utilization involve modifying the codebook lookup mechanism, including the improvement of distance calculation of quantitative mapping [23], and development of learning strategies of codebook [24].  
Overall, the above studies mainly focus on quantitative mapping performance, based on which the reconstruction and generation tasks are carried out.

For semantic communication, another important aspect is to ensure the effective transmission of codewords after quantitative mapping.
The index sequence of codewords is inevitably influenced by wireless fading channel and additive noise, which results in incorrect decoding. 
In [17], the proposed scheme adopts the perfect channel state information (CSI) and channel coding to ensure the accuracy of index transmission. 
In [25], established on a discrete memoryless channel (DMC), the channel-aware vector quantization (CAVQ) method designs a loss function that explicitly incorporates the channel transition probability and aggregates the codewords with high error probabilities in the semantic space to minimize the transmission error. 
In [26], the parallel binary symmetric channels (BSCs) with trainable bit-flip probabilities model binary output transmission and reception, where the multiple VQ codebooks and their associated bit-flip probabilities are jointly optimized.
These studies provide insights into VQ-enabled DSC system with simplified channel models.
Recent studies have investigated the impacts of multipath fading channels on semantic communication, such as the semantic importance-aware reordering-enhanced semantic communication system (SIARE-SC) based on estimated channel [27] and  channel denoising diffusion model (CDDM) for semantic communication [28].  
However, they haven't integrated with the VQ-enable architecture.

Motivated by these issues, this paper aims to resolve three pivotal questions:

\textit{Question1: How to design an efficient VQ-enabled DSC scheme that is compatible with practical communication systems?} 

\textit{Question2: How to train a VQ codebook for efficient quantitative mapping related to semantic features?}

\textit{Question3: How to achieve end-to-end optimization of digital semantic encoder, VQ codebook, and decoder in complex communication scenarios?}

\begin{figure*}[htbp]
\centering
\resizebox{1\textwidth}{0.20\textheight}{\includegraphics{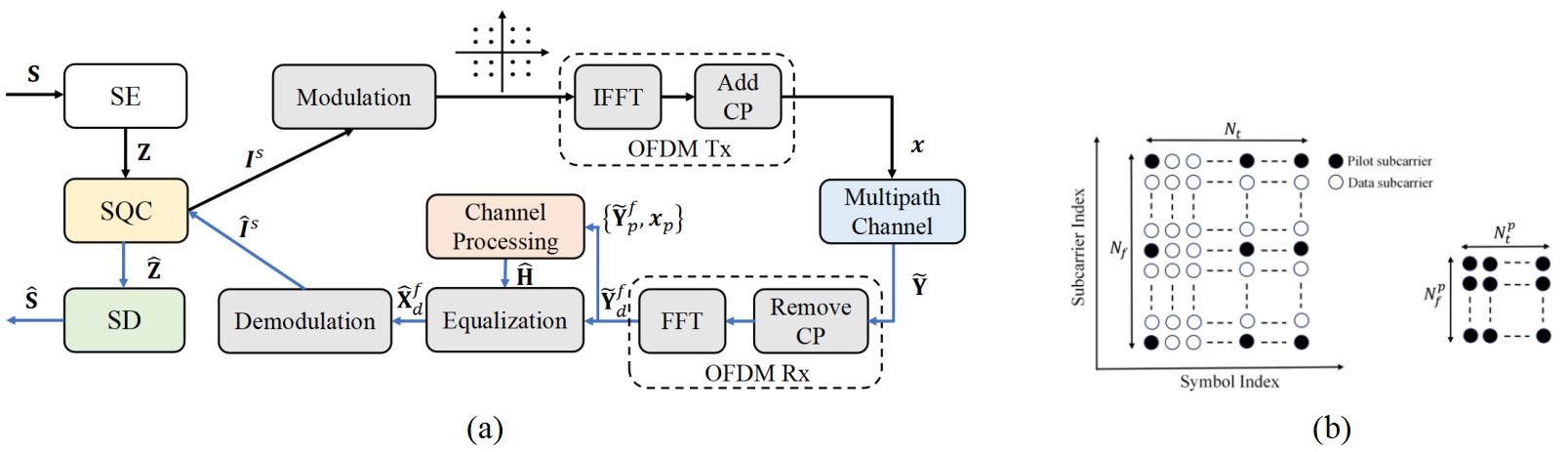}}
\caption{Architecture of VQ-DSC-R system with OFDM transmission. (a) Overall system architecture; (b) Illustration of time-frequency grids of OFDM symbols.}
\label{fig1}
\end{figure*}

In this paper, we develop a robust VQ-enabled DSC (VQ-DSC-R) system with OFDM transmission. Especially, the VQ-DSC-R scheme is optimized over multipath fading channel and noise in OFDM-based system, where the issues of semantic feature extraction/reconstruction, quantification and transmission are simultaneously considered. Our key contributions are outlined as follows:
\begin{itemize}
\item[]
\hspace{-0.5cm} $\bullet$
\hspace{0.1cm}\textit{Multi-Stage Vector Quantization Framework for Semantic Feature Extraction and Quantitative Mapping: } 
We propose a cascaded architecture that consists of a Swin Transformer-based backbone [29] for hierarchical semantic feature extraction, followed by VQ modules that project these features into vectors in a shared semantic quantized codebook
(SQC) of transceiver.
The trainable SQC enhances the interpretability of semantic communication.
The index sequence related to extracted features is transmitted using adaptive modulation over a practical OFDM system, resilient to multipath fading and additive noise.
This addresses the \textit{Question 1}.

\hspace{-0.5cm} $\bullet$
\hspace{0.1cm}\textit{SQC Update Scheme for Efficient Quantitative Mapping of Semantic Features:} 
We develop a SQC update scheme, named the differentiable vector quantization with adaptive noise variance (ANDVQ). Specifically, it introduces the K-nearest neighbor adaptive noise variance to dynamically adjust quantization process. Meanwhile, exponential moving averages (EMA) mechanism is integrated to stabilize SQC training. In this way, the issues of gradient collapse and codebook collapse are solved, thereby mitigating quantization errors related to semantic features. This addresses the \textit{Question 2}.

\hspace{-0.5cm} $\bullet$
\hspace{0.1cm}\textit{Channel Distortion Reduction and Training Strategy for VQ-DSC-R:}
We first analyze the impacts of CSI accuracy on end-to-end training process. To mitigate it, the conditional diffusion model (CDM)-based scheme for CSI refinement is proposed, which can generate accurate CSI data within few reverse sampling.
Additionally, the signal-to-noise ratio (SNR) adaptive module is designed and merged into semantic decoder to counter different SNR conditions. Finally, a three-stage training strategy is developed to achieve optimal overall performance of VQ-DSC-R system. This addresses the \textit{Question 3}.

\hspace{-0.5cm} $\bullet$
\hspace{0.1cm}\textit{Experimental Validation:}
To verify the performance of the proposed VQ-DSC-R system, we have conducted a large number of experimental validations under the 3GPP standard channel model and OFDM transmission, and compared with those of popular SQC update schemes and channel estimation schemes. Utilizing the transmission index sequence of codebook related to quantitative mapping of semantic features, the results show that the proposed scheme can achieve effective transmission with an extremely high compression rate in terms of peak SNR (PSNR) and multiscale structural similarity (MS-SSIM) metrics. Meanwhile, the results comprehensively present the influence of multipath fading channels and SNRs on overall performance. 

\end{itemize}

The remainder of this paper is organized as follows. Section II provides an overview of the system model and problem formulation. Section III elaborates the proposed VQ-DSC-R system, including the framework design and its system optimization study.
Section IV presents experimental results that validate the performance of the proposed scheme. Section V concludes the paper.

\textit{Notation:} Boldface small letters denote vectors and boldface capital letters denote matrices. 
$\mathbb{R}$, $\mathbb{Z}$ and $\mathbb{C}$ denote the real number field, the integer number field, and the complex number field, respectively. 
$[\cdot]^{T}$ denotes matrix transpose. 
$\|\cdot\|_{2}$ denotes the Euclidean norm.
$\ast$ denotes the convolution operation. 
$\mathcal{CN}(\cdot|\mu,\Gamma)$ denotes the complex Gaussian distribution with mean $\mu$ and covariance $\Gamma$.
$\mathbb{E}_{p(x)}[\cdot]$ denotes the expectation with respect to the distribution $p(x)$.

\section{System model and problem formulation}

To address the limitations of conventional SSCC, this work proposes a digital DJSCC framework that synergizes DNNs with digital communication principles.
Fig. 1(a) illustrates the proposed VQ-DSC-R system with OFDM transmission, where the black arrow and blue arrow represent the data stream at the transmitting side and receiving side, respectively. The semantic encoder (SE) first extracts key features of the source, which are mapped into discrete vectors within a shared SQC. Then the corresponding index sequence of those vectors are transmitted to the receiver through wireless channel, rather than the features themselves, substantially improving efficiency. The semantic decoder (SD) reconstructs the original content with the aid of SQC.
Note that the SQC is trained along with the SE and SD. 

\subsection{System Model}
Without loss of generality, the source dataset consists of images. Let the image dataset be $\bm{\chi}$, with each image  $\mathbf{S} \in \mathbb{R}^{H\times W \times O} \subset \bm{\chi}$, where $H$, $W$ and $O$ are the height, width, and the channel size of each image, respectively. The trainable SQC is given by $\mathbf{\bm{C}} = \left[\bm{c}_{1},..., \bm{c}_{k},..., \bm{c}_{K} \right]^{T} \in \mathbb{R}^{K \times N}$, where $K$ is the codebook size, and each element $\bm{c}_{k} \in \mathbb{R}^{N}$ is called a semantic codeword with dimension size $N$ and index $k$. The SQC serves as a consistently shared semantic knowledge base for the transmitter and the receiver. 
The SE and SD are combined with the SQC that jointly completes the semantic extraction, quantification and reconstruction of images. 
The extracted semantic feature vector $\mathbf{\bm{Z}} = \left[\bm{z}_{1},..., \bm{z}_{k},..., \bm{z}_{K} \right]^{T} \in \mathbb{R}^{K \times M}$ of the input image can be expressed as 
\begin{equation}
\mathbf{\bm{Z}} = f_{SE}\left(\mathbf{\bm{S}}, \bm{\alpha} \right),
\setcounter{equation}{1}
\end{equation}
where $f_{SE}\left(\cdot, \bm{\alpha}\right)$ denotes the \textit{SE} module built on DNN with the parameter set $ \bm{\alpha}$.

Then, the semantic feature element $\bm{z}_{k}$ is mapped into the SQC by the nearest-neighbor operation, i.e., $q\left(\bm{z}_{k} \right)$, and the quantized feature can be expressed as 
\begin{equation}
\hat{\bm{z}}_{k}=q\left(\bm{z}_{k} \right) \triangleq \operatorname*{argmin}_{\bm{c}_{j}} \left\| \bm{z}_{k} - \bm{c}_{j} \right\|_{2}, j=1,2,...,K.
\setcounter{equation}{2}
\end{equation}

After all the semantic features are mapped into the discrete space, the corresponding index sequence of semantic features of $\mathbf{\bm{Z}}$ is obtained, which is denoted as $\bm{I}^{s} \in \mathbb{Z}^{M}$. SQC-enabled quantization operation transforms the float-point semantics to discrete integer indices through a trained codebook mapping, effectively establishing a discrete semantic space that approximates the original continuous distribution of semantic features. By using the \textit{Modulation} module, $\bm{I}^{s}$ is turned into a symbol vector $\bm{x}\in \mathbb{C}^{B}$ for transmission, where $B$ denotes the length of modulated symbols.

The DJSCC framework integrated with an OFDM-based system is developed with considering the multipath fading channel and additive channel noise. For each physical resource block (PRB) defined in digital communication systems (such as the 3GPP standard [30]), it consists of $N_{f}$ subcarriers of each OFDM symbol and $N_{t}$ OFDM symbols. To improve spectral efficiency, the uniform grid-type pilot-aided channel estimation where the $N_{f}^{p}$ subcarriers and $N_{t}^{p}$ symbols are used for pilot transmission, as shown in Fig. 1(b). Assume the set of subcarriers that carry pilots $\bm{p}_{f}=\left\{p_{1}^{f}, p_{2}^{f},...,p_{N_{f}^{p}}^{f} \right\}$ under $\bm{p}_{f} \subset \left\{1,2,...,N_{f} \right\}$, and the set of symbols that carry pilots $\bm{p}_{t}=\left\{p_{1}^{t}, p_{2}^{t},...,p_{N_{t}^{p}}^{t} \right\}$ under $\bm{p}_{t} \subset \left\{1,2,...,N_{t} \right\}$. The pilot symbols $\bm{x}_{p} \in \mathbb{C}^{N_{p}}$ are known to both the transmitter and receiver, where $N_{p} = N_{f}^{p} + N_{t}^{p}$. The symbol vector $\bm{x}$ is reshaped to $\bm{x} \in \mathbb{C}^{\left(N_{f}\times N_{s} - N_{p}\right)}$, where $N_{s}$ denotes the number of OFDM information symbols. In the case, the $N_{s} = \lceil \frac{B }{N_{f} \times N_{t} } \rceil$, where $\lceil \cdot \rceil$ denotes the ceiling operation. Next, we apply the inverse discrete Fourier transform (IDFT) to each OFDM symbol and append the cyclic-prefix (CP). 
Assume that the transmitter and receiver are synchronized with the cyclic-prefix length $N_{cp}$.
The transmitted signal $\tilde{\mathbf{\bm{X}}} \in \mathbb{C}^{\left(\left(N_{f} + N_{cp} \right)\times N_{s} \right) \times N_{t}} $ propagates through the multipath fading channel, and the received signal can be expressed as 
\begin{equation}
\tilde{\mathbf{\bm{Y}}} = \bm{h} \ast \tilde{\mathbf{\bm{X}}} + \mathbf{\bm{W}},
\setcounter{equation}{3}
\end{equation}
where the $\ast$ denotes the convolution operator, $\bm{h} \in \mathbb{C}^{L}$ denotes the channel impulse response with $L$ denoting the number of delays, $\mathbf{\bm{W}}$ denotes the additive Gaussian noise that satisfies $\mathcal{CN}\left(0, \sigma^{2}\mathbf{I} \right)$, and $\sigma^{2}$ denotes noise power. 

\begin{figure*}[htbp]
\centering
\resizebox{1\textwidth}{0.42\textheight}{\includegraphics{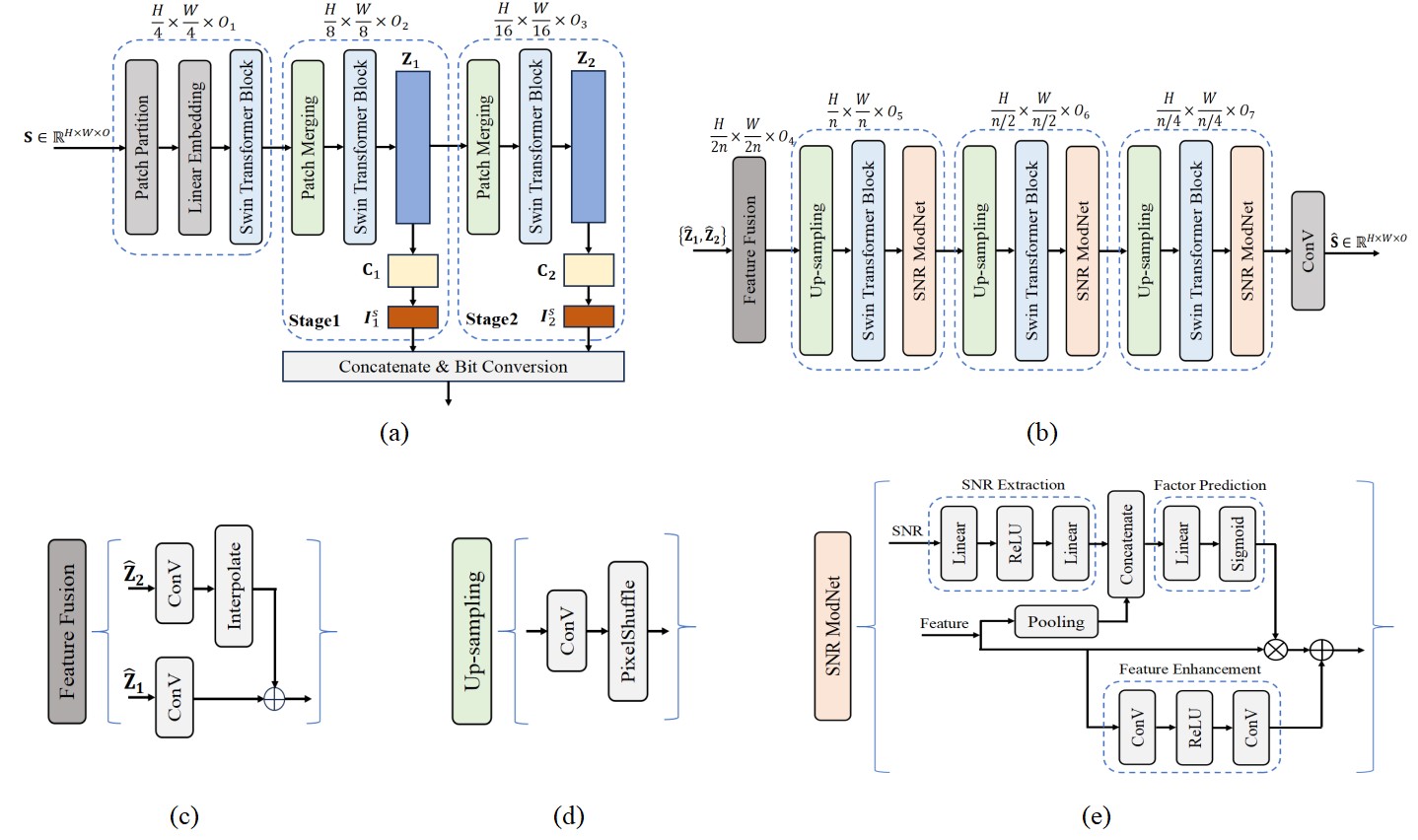}}
\caption{Architectures of SQC-SE and SQC-SD for DSC. (a) Architecture of SQC-SE; (b) Architecture of SQC-SD; (c) 
\textit{Feature Fusion} layer; (d) \textit{Up-sampling} layer; (e) \textit{SNR ModNet} layer.}
\label{fig1}
\end{figure*}

After removing the CP, the frequency signals $\tilde{\mathbf{\bm{Y}}}^{f} = \left\{\tilde{\mathbf{\bm{Y}}}^{f}_{p}, \tilde{\mathbf{\bm{Y}}}^{f}_{d} \right\} $ can be obtained by discrete Fourier transform (DFT), where $\tilde{\mathbf{\bm{Y}}}^{f}_{p}$ denotes the pilot signals and $\tilde{\mathbf{\bm{Y}}}^{f}_{d}$ denotes the data signals. By using $\tilde{\mathbf{\bm{Y}}}^{f}_{p}$ and pre-inserted pilots $\bm{x}_{p}$, the \textit{Channel Processing} module is developed to obtain the whole frequency CSI of PRBs. To mitigate the bad impacts of inaccurate estimated CSI on index transmission and end-to-end training, we propose a \textit{Channel Processing} module for estimated CSI refinement. Specifically, the conditional diffusion model (CDM)-based CSI reconstruction scheme is developed, which can provide accurate CSI with very few sampling steps. 
The estimated frequency channel can be expressed as
\begin{equation}
\hat{\mathbf{\bm{H}}} = f_{cp}\left(\left\{\tilde{\mathbf{\bm{Y}}}^{f}_{p}, \bm{x}_{p}\right\}, \bm{\gamma} \right),
\setcounter{equation}{4}
\end{equation}
where $f_{cp}\left(\cdot, \bm{\gamma} \right)$ denotes the \textit{Channel Processing} module built on DNN with the parameter set $\bm{\gamma}$.

The \textit{Channel Equalization} module is used to recover data symbols $\hat{\mathbf{\bm{X}}}_{d}^{f}$ by using $\hat{\mathbf{\bm{Y}}}_{d}^{f}$ and $\hat{\mathbf{\bm{H}}}$. The indices $\hat{\bm{I}}^{s}$ is demodulated from $\hat{\mathbf{\bm{X}}}_{d}^{f}$ by \textit{Demodulation} module, based on which the quantitated vectors $\hat{\mathbf{\bm{Z}}}$ mapped by SQC can be obtained. The original image can be finally reconstructed as 
\begin{equation}
\hat{\mathbf{\bm{S}}} = f_{SD}\left(\hat{\mathbf{\bm{Z}}}, \bm{\beta} \right), 
\setcounter{equation}{5}
\end{equation}
where $ f_{SD}\left(\cdot, \bm{\beta} \right)$ denotes the \textit{SD} module built on DNN with the parameter set $\bm{\beta}$.

\subsection{Problem Formulation}

To effectively characterize continuous semantic features by limited codebook capacity, the DNN for semantic feature extraction integrated with quantization mapping requires meticulous architectural design. 
On the other hand, the index transmission of vectors in SQC is susceptible to the effects of nultipath fading channel and additive channel noise, which therefore significantly reduces the overall decoding performance. Considering the constraints of SQC and index transmission in wireless communication scenarios, the semantic encoder-decoder parameter set $\left\{\bm{\alpha}, \bm{\beta} \right\}$, channel processing parameter set $\bm{\gamma}$ and trainable codebook $\mathbf{\bm{C}}$ are optimized by minimizing the expectation as follows:
\begin{equation}
\left(\bm{\alpha}^{*}, \bm{\beta}^{*}, \bm{\gamma}^{*}, \mathbf{\bm{C}}^{*}\right) = \operatorname*{argmin}_{\bm{\alpha}, \bm{\beta}, \bm{\gamma}, \mathbf{\bm{C}}} \mathbb{E}_{p(\sigma)}\mathbb{E}_{p(\bm{h})}\mathbb{E}_{p(\mathbf{S}, \hat{\mathbf{S}})}\left[d\left(\mathbf{S}, \hat{\mathbf{S}} \right)  \right],
\setcounter{equation}{6}
\end{equation}
where $\bm{\alpha}^{*}$ is the optimal parameter set of SQC-SE, $\bm{\beta}^{*}$ is the optimal parameter set of SQC-SD, $\bm{\gamma}^{*}$ is the optimal parameter set of \textit{Channel Processing} module, $\mathbf{\bm{C}}^{*}$ is the optimal codebook, $p(\sigma)$ denotes the probability distribution of channel noise, $p(\bm{h})$ denotes the probability distribution of multipath fading channel, $p(\mathbf{S}, \hat{\mathbf{S}})$ denotes the joint probability distribution of the original image $\mathbf{S}$ and the reconstructed image $\hat{\mathbf{S}}$, and $d\left(\mathbf{S}, \hat{\mathbf{S}} \right) $ denotes the distortion between $\mathbf{S}$ and $\hat{\mathbf{S}}$, namely $d\left(\mathbf{S}, \hat{\mathbf{S}} \right) = \frac{1}{n}\sum_{i=1}^{n}\left(s_{i} - \hat{s}_{i}\right)^{2}$, where $s_{i}$ and $\hat{s}_{i}$ represent the intensity of the color component of each pixel corresponding to $\mathbf{S}$ and $\hat{\mathbf{S}}$, respectively. 
The subsequent work is to develop a SQC integrated with DSC system to minimize source reconstruction errors with considering the aforementioned \textit{Questions 1-3}.

\section{Proposed VQ-DSC-R System}
In this section, we provide the detailed description of VQ-DSC-R system, which mainly includes the semantic feature extraction/reconstruction, quantitative mapping, and transmission with considering multipath fading channel and noise. 

\subsection{VQ-enabled Semantic Feature Extraction, Quantization, Transmission and Reconstruction}
This subsection details our proposed architectures of SQC-SE and SQC-SD for DSC, where the Swin Transformer-based backbone is integrated with VQ mechanism for joint semantic feature extraction, quantization and reconstruction. For the SQC-SE, we design a two-stage architecture to extract semantic features and mitigate quantization errors, where each stage aligns with distinct feature dimensions and dedicated quantized codebooks. For the SQC-SD, the features are gradually reconstructed, and the SNR adaptive module is designed to reduce the impact of different SNR conditions on index transmission.
The designed SNR adaptive module is not deployed at the transmitter side, which is different from other works [31]. 
This is because the model with dual-end deployment is not applicable for 
the three-stage training strategy (described in Section III-D).

An overview of the proposed architectures of SQC-SE and SQC-SD for wireless image transmission is presented in Fig. 2(a) and Fig. 2(b), respectively. As to the SQC-SE, the source S is first partitioned into $\frac{H}{n} \times \frac{W}{n}$ non-overlapping patches by \textit{Patch Partition} layer, which are regarded as tokens and arranged in a sequence by following a left-to-right and top-to-bottom order. In this case, $n$ denotes the integer factor of down-sampling. Then it is followed by \textit{Linear Embedding} layer to map the patches into a semantic latent representation with size $\frac{H}{n} \times \frac{W}{n} \times O_{1}$. Subsequently, a \textit{Swin Transformer Block} layer is applied to these features. Within Swin Transformer framework, it employs two key attention mechanisms, namely the window multi-head self-attention (W-MSA) and shifted window multi-head self-attention (SW-MSA). 
The core idea is brief summarized: 1) W-MSA restricts self-attention computation to non-overlapping local windows, dramatically reducing computational complexity compared to global attention; 2) To overcome the limited cross-window connectivity, SW-MSA applies self-attention within windows shifted by half the window size relative to W-MSA. It enables communication between adjacent windows from the previous layer, which therefore allows the model to capture long-range dependencies. For a detailed explanation of W-MSA and SW-MSA, please refer to [29].

To build a hierarchical representation with SQC, each stage incorporates a \textit{Patch Merging} layer followed by \textit{Swin Transformer Block} layer. Taking “Stage 1” as an example, neighboring embeddings are merged by a \textit{Patch Merging} layer, resulting in concatenated embeddings of size $O_{1}$ upgrading to size $O_{2}$. Subsequently, $\frac{H}{2n} \times \frac{W}{2n}$ patch embedding tokens with higher-resolution are fed into \textit{Swin Transformer Block} layer. In this way, the complex details of embedded features can be learned by capturing long-range dependencies and exploiting global information. As to the extracted semantic feature $\mathbf{Z}_{1}$, it mapped into the codebook $\mathbf{C}_{1}$ based on Eq. (2), and the corresponding index sequence is obtained, namely $\hat{\bm{I}}_{1}^{S} \in \mathbb{R}^{M_{1}}$. Similarly, $\hat{\bm{I}}_{2}^{S} \in \mathbb{R}^{M_{2}}$ can be obtained in “Stage 2”. These indices are cascaded to form an index vector and turned into a bit sequence for transmission by \textit{Concatenate \& Bit Conversion} layer. Given the critical impact of quantized codebook expressiveness on semantic communication fidelity, Section II-C details our proposed SQC update scheme.

As to the SQC-SD shown as Fig. 2(b), the \textit{Feature Fusion} layer first aligns multi-scale feature $\hat{\mathbf{Z}}_{1}$ and $\hat{\mathbf{Z}}_{2}$ via combined \textit{ConV} and \textit{Interpolation}, and then synthesizes them through element-wise summation. The architecture of \textit{Feature Fusion} layer is shown in Fig. 2(c), where the \textit{ConV} denotes the convolutional operator and \textit{Interpolation} denotes the nearest neighbor interpolation operator. The \textit{Up-sampling} layer, \textit{Swin Transformer Block} layer and \textit{SNR ModNet} layer are cascaded to gradually reconstruct semantic features as well as adapt to different channel conditions. By \textit{Up-sampling} layer, the dimension of features is Upsampled by a factor of 2, and the embedding of size $O_{4}$ is reduced to size $O_{5}$. The architecture of Up-sampling layer is shown in Fig. 2(d), where the \textit{PixelShuffle} denotes the operator that rearranges elements in a tensor according to an upscaling factor [32]. Subsequently, $\frac{H}{n} \times \frac{W}{n}$ patch embedding tokens with lower-resolution are fed into \textit{Swin Transformer Block} layer for feature reconstruction, and then a \textit{SNR ModNet} layer is followed to improve the channel adaptive ability. 

As shown in Fig. 2(e), the \textit{SNR ModNet} layer is designed to improve its adaptive ability to changing channel conditions.
By leveraging channel-wise soft attention mechanism, it dynamically generates scaling parameters conditioned on the instantaneous SNR. Firstly, the SNR is input into SNR Extraction module to balance the wide SNR range. Specifically, the SNR Extraction module consists of \textit{Linear} operator and \textit{ReLU} operator. Secondly, the feature is processed by using a global average pooling function, denoted as Pooling operator, which is followed by the Factor Prediction module to predict the scaling factor based on the concatenated feature and SNR information. The Factor Prediction module consists of \textit{Linear} operator and \textit{Sigmoid} operator. In addition, the Feature Enhancement module is adopted as a residual link for feature enhancement, which consists of \textit{ConV} and \textit{ReLU} operators. Finally, the features that have been superimposed with SNR information are obtained.

\subsection{Proposed ANDVQ Scheme for SQC Update}

Joint optimization of the SQC with SE and SD requires addressing two fundamental issues, as mentioned in Section I-B.
The proposed ANDVQ scheme is inspired by noise substitution in vector quantization (NSVQ) scheme proposed in [22]. The main idea of NSVQ is to simulate the VQ distortion by adding Gaussin noise $\mathcal{N}\left(0, 1\right)$ to the latent vector, which makes the quantization operation be differentiable in terms of the latent feature and selected codeword. 
However, as the noise is sampled randomly, it introduces arbitrary quantization errors and challenges for the codebook to converge to optimum location. To deal with it, 
the ANDVQ scheme introduces the K-nearest neighbor adaptive noise variance to dynamically adjust the quantization process through local codebook statistics. 
Specifically, it approximates the quantized vector $\bm{z}_{q}$ as the addition of the quantization error $\bm{\epsilon}_{q}$ to the input vector $\bm{z}$, i.e., $\bm{z}_{q} = \bm{z} + \bm{\epsilon}_{q}$, which can be calculated as 
\begin{equation}
\bm{z}_{q} = \bm{z} + \left\|\bm{d}_{avg} \right\|_{2} \cdot \mathrm{sg}\left[\frac{\bm{v}_{d} }{\left\|\bm{v}_{d} \right\|_{2}   } \right],
\setcounter{equation}{7}
\end{equation}
\begin{equation}
\bm{v}_{d} = \bm{d}_{avg} + \bm{v}_{n},
\setcounter{equation}{8}
\end{equation}
where $\mathrm{sg}[\cdot]$ denotes the stop gradient operation, $\bm{v}_{d}$ denotes the constructed direction vector, in which $\bm{d}_{avg}$ provides the deterministic direction guidance aligned with the nearest codeword, and $\bm{v}_{n}$ provides random exploration. In this way, $\bm{v}_{d}$ can balance quantization accuracy and model generalization ability. 
Specifically, $\bm{d}_{avg}$ is calculated based on the average offset direction of all $K_{c}$ nearest neighbors relative to $\bm{z}$, which can be expressed as 
\begin{equation}
\bm{d}_{avg} = \frac{1}{K_{c}}\sum_{\bm{c}_{i_{k}} \subseteq \mathbf{C}_{K_{c}-NN}}\left(\bm{z} - \bm{c}_{i_{k}} \right),
\setcounter{equation}{9}
\end{equation}
where $\mathbf{C}_{K_{c}-NN}=\left\{\bm{c}_{i_{1}}, \bm{c}_{i_{2}},...,\bm{c}_{i_{k}}  \right\} \subseteq \mathbf{C}$ denotes the $K_{c}$ nearest neighbors with respect to $\bm{z}$, which is calculated based on Euclidean distance, i.e.,
\begin{equation}
\mathbf{C}_{K_{c}-NN} = \operatorname*{argmin}_{\mathbf{C}_s \subseteq \mathbf{C}, \left| \mathbf{C}_s \right|=K_{c}}\sum_{\bm{c}_{i_{k}} \subseteq \mathbf{C}_s} \left\|\bm{z} - \bm{c}_{i_{k}}  \right\|_{2}.
\setcounter{equation}{10}
\end{equation}
Meanwhile, $\bm{v}_{n} $ denotes the adaptive noise adapted to local density in terms of $\bm{z}$, which satisfies $\mathcal{N}\left(0, \sigma_{q}^{2} \right)$. 
The $\sigma_{q}$ is calculated by the arithmetic mean of $K_{c}$ nearest neighbor distances in terms of $\bm{z}$, which is calculated as
\begin{equation}
\sigma_{q} = \frac{1}{K_{c}} \sum_{\bm{c}_{i_{k}} \subseteq \mathbf{C}_{K_{c}-NN}}\left\| \bm{z} - \bm{c}_{i_{k}} \right\|_{2}. 
\setcounter{equation}{11}
\end{equation}

By using Eq. (7) during training, $\bm{z}_{q}$ is differentiable function of feature $\bm{z}$ and codeword $\bm{c}_{k}$, 
and the gradients can be calculated as 
\begin{equation}
\frac{\partial \bm{z}_{q} }{\partial \bm{z}} = 1 - \mathrm{sg}\left[\frac{\bm{d}_{avg} }{\left\|\bm{d}_{avg}  \right\|_{2} } \right] \cdot \frac{\bm{d}_{avg} }{\left\|\bm{d}_{avg}  \right\|_{2} },
\setcounter{equation}{12}
\end{equation}
\begin{equation}
\frac{\partial \bm{z}_{q} }{\partial \bm{c}_{k}} =  \mathrm{sg}\left[\frac{\bm{d}_{avg} }{\left\|\bm{d}_{avg}  \right\|_{2} } \right] \cdot \frac{\bm{d}_{avg} }{\left\|\bm{d}_{avg}  \right\|_{2} }.
\setcounter{equation}{13}
\end{equation}

During $e$-th training epoch, the cumulative number of semantic features assigned to the quantized vector $\bm{c}_{k}$ in SQC can be calculated as 
\begin{equation}
\varphi_{k}^{(e)} = \gamma_{0}\varphi_{k}^{(e-1)} + (1-\gamma_{0})\psi_{k}^{(e)}, 
\setcounter{equation}{14}
\end{equation}
where $\varphi_{k}^{(e)}$ denotes the number of semantic features mapped to $\bm{c}_{k}$ in the $e$-th training epoch, $\psi_{k}^{(e)}$ denotes the current frequency usage of $\bm{c}_{k}$ in the codebook, and $\gamma_{0} \in[0, 1)$ denotes a hyper-parameter that represents the decay rate.

The cumulative sum of the features assigned to the quantized vector $\bm{c}_{k}$ can be calculated as 
\begin{equation}
{\Phi}_{k}^{(e)} = \gamma_{0}\Phi_{k}^{(e-1)} + (1-\gamma_{0})\sum_{\bm{c}_{k} \leftarrow \bm{z}_{q}}\bm{z}_{q}.
\setcounter{equation}{15}
\end{equation}
The codebook in the $e$-th epoch can be updated as 
\begin{equation}
\bm{c}_{k}^{(e)} = \frac{\Phi_{k}^{(e)}}{\varphi_{k}^{(e)} + \epsilon}, n=1,2,...,N,
\setcounter{equation}{16}
\end{equation}
where $\epsilon$ is a hyperparameter that prevents division by zero.

\subsection{Proposed CDM-based Scheme for CSI Refinement}
The impacts of multipath fading channel need to be mitigated by channel equalization in traditional communication systems. Next, we will analyze the role of CSI in our end-to-end training system. If the perfect frequency channel $\mathbf{H}_{0}$ is known, the received symbol can be processed by 
\begin{equation}
\tilde{\mathbf{X}}^{f} = \left(\mathbf{H}_{0}^{H}\mathbf{H}_{0}\right)^{-1}\mathbf{H}_{0}^{H}\tilde{\mathbf{\bm{Y}}}^{f} = \mathbf{X}^{f} + \tilde{\mathbf{W}}, 
\setcounter{equation}{17}
\end{equation}
where $\mathbf{X}^{f}$ denotes the original transmitted signal, and $\tilde{\mathbf{W}} = \left(\mathbf{H}_{0}^{H}\mathbf{H}_{0} \right)^{-1}\mathbf{H}_{0}^{H}\mathbf{W}$. 

From Eq. (17), if the channel is perfectly estimated,  the channel effect is transferred from multiplicative interference to additive noise $\tilde{\mathbf{W}}$, which provides the possibility of stable back-propagation as well as the stronger capability of network representation. In this case, the back-propagation and forward-propagation can be performed by setting $\mathbf{H}_{0}$ being equal to the identity matrix.
However, the perfect CSI cannot be obtained in practice. 
Pilot-aided channel estimation includes two steps as follows:  
the CSI at pilot positions is first estimated, and then the interpolation operation is adopted to obtain the whole CSI of PRBs. Specifically, the rough CSI at pilot positions is estimated by the channel estimator with few inserted pilot signals, which can be expressed as 
\begin{equation}
\mathbf{H}_{p}^{r} = \mathbf{H}_{p} + \mathbf{W}\mathbf{X}_{p}^{H},
\setcounter{equation}{18}
\end{equation}
where $\mathbf{H}_{p}$ denotes the perfect frequency channel at pilot positions, and $\mathbf{X}_{p}$ denotes the frequency pilot signals. 
Then the whole CSI (including pilot channel and data channel) can be obtained by 
\begin{equation}
\mathbf{H}^{r} = f_{interp}\left(\mathbf{H}_{p}^{r} \right),
\setcounter{equation}{19}
\end{equation}
where $f_{interp}\left(\cdot \right)$ denotes the interpolation function. 

From (19), $\mathbf{H}^{r}$ is inaccurate due to channel noise and interpolation error. Next, we develop a CDM-based method to refine $\mathbf{H}^{r}$. 
Specifically, in the forward diffusion process, it gradually adds noise to the training data. 
Then, in the reverse sampling process, it learns to recover the true data from the perturbed data. 
By adding well-planed noise, our proposed scheme can significantly reduce the number of reverse sampling. 

\textit{ 1) Forward Diffusion Process:}  

Assume that the real part and imaginary part of $\mathbf{H}_{0}\left(\mathbf{H}^{r} \right)$ are extracted to form a real vector $\bm{h}_{0}\left(\bm{h}_{0}^{r} \right)$. 
In this case, we regard $\bm{h}_{0}$ as the label data, and gradually add noise conditioned on $\bm{h}_{0}^{r}$ and SNR value $v_{s}$. The transformed distribution can be modeled as
\begin{equation}
q\left(\bm{h}_{t} | \bm{h}_{t-1}, \bm{h}_{0}^{r}, v_{s} \right)\sim \mathcal{N}\left(\bm{h}_{t} | \bm{h}_{t-1}+\alpha_{t}\Delta_{h}, \kappa_{s}^{2}\alpha_{t}\mathbf{I} \right),
\setcounter{equation}{20}
\end{equation}
where $\left\{\eta_{t} \right\}_{t=1}^{T}$ denotes the sequence of transformation with satisfying $\eta_{1}\to 0$ and $\eta_{T}\to 1$, $\Delta_{h}$ denotes the difference between $\bm{h}_{0}^{r}$ and $\bm{h}_{0}$, i.e., $\Delta_{h}=\bm{h}_{0}^{r}-\bm{h}_{0}$, $\kappa_{s}$ denotes noise adjustment parameter, $\alpha_{t}=\eta_{1}$ under $t=1$ and $\alpha_{t}=\eta_{t} - \eta_{t-1} $ under $t \ge 1$. 

By using reparameterization trick, $\bm{h}_{t}$ can be obtained by 
\begin{equation}
\bm{h}_{t} = \bm{h}_{t-1} + \alpha_{t} \Delta_{h} + \kappa_{s}\sqrt{\alpha_{t}}\bm{\xi}_{t}, \bm{\xi}_{t} \sim \mathcal{N}\left(0,1\right).
\setcounter{equation}{21}
\end{equation}
Through recursive derivation, $\bm{h}_{t}$ can be further rewritten as 
\begin{equation}
\bm{h}_{t} = \bm{h}_{0} + \eta_{t}\Delta_{h} + \kappa_{s}\sqrt{\alpha_{t}}\bm{\xi}_{t}.
\setcounter{equation}{22}
\end{equation}
So, $\bm{h}_{t}$ satisfies the Gaussian distribution as follows
\begin{equation}
q\left(\bm{h}_{t} | \bm{h}_{0}, \bm{h}_{0}^{r}, v_{s} \right) \sim \mathcal{N}\left(\bm{h}_{t} | \bm{h}_{0} + \eta_{t} \Delta_{h}, \kappa_{s}^{2}\eta_{t}\mathbf{I} \right).
\setcounter{equation}{23}
\end{equation}

From Eq. (23), it indicates that the added noise level of $\bm{h}_{t}$ is determined by $\kappa_{s}$ and $\sqrt{\eta_{t}}$. By properly planning the method of adding noise, it is possible to cover the range from low noise to high noise, which benefits reducing the number of processing steps required for adding noise and removing noise to generate the actual channel data distribution. The $\kappa_{s}$ and $\sqrt{\eta_{t}}$ are calculated as follows:
\begin{equation}
\sqrt{\eta_{t}} = \sqrt{\eta_{1}} \times \lambda^{\beta_{t}},
\setcounter{equation}{24}
\end{equation}
\begin{equation}
\kappa_{s} = \kappa_{0} e^{-\frac{v_{s}}{10}},
\setcounter{equation}{25}
\end{equation}
where $\lambda_{0}=\frac{1}{e^{2(T-1)\log\frac{\eta_{T} }{\eta_{1} }}}$, $\beta_{t}=\left(\frac{t-1}{T-1}\right)^{\rho_{0}}\times\left(T-1\right)$, $\kappa_{0}$ and $\rho_{0}$ are hyper-parameters. Note that the calculation method of $\sqrt{\eta_{t}}$ is similar to [33]. 

\textit{2) Reverse Sampling Process:}  

The reverse sampling process intends to sample the original data distribution from $\bm{h}_{0}^{r}$. The equivalent posterior distribution estimation can be expressed as 
\begin{align}
p\left(\bm{h}_{0} | \bm{h}_{0}^{r}, v_{s} \right) & =  \nonumber\\
& \hspace{-1cm} \int p\left(\bm{h}_{T} | \bm{h}_{0}^{r}, v_{s}  \right) \prod_{t=1}^{T}p_{\bm{\gamma}}\left(\bm{h}_{t-1} | \bm{h}_{t}, \bm{h}_{0}^{r}, v_{s} \right)d\bm{h}_{1:T}.
\setcounter{equation}{25}
\label{eq18}
\end{align}
with
\begin{equation}
p\left(\bm{h}_{T} | \bm{h}_{0}^{r}, v_{s} \right) \sim \mathcal{N}\left(\bm{h}_{T} |  \bm{h}_{0}^{r}, v_{s} \right),
\setcounter{equation}{27}
\end{equation}
\begin{align}
p_{\bm{\gamma}}\left(\bm{h}_{t-1} | \bm{h}_{t}, \bm{h}_{0}^{r}, v_{s} \right) \sim &  \nonumber\\
& \hspace{-2cm} \mathcal{N}\left(\bm{h}_{t-1} | \bm{\mu}_{\bm{\gamma}}\left(\bm{h}_{t}, \bm{h}_{0}^{r}, v_{s} \right), \bm{\Sigma}_{\bm{\gamma}}\left(\bm{h}_{t}, \bm{h}_{0}^{r}, v_{s}  \right)   \right).
\setcounter{equation}{27}
\label{eq18}
\end{align}
where $p_{\bm{\gamma}}\left(\cdot \right)$ denotes the inverse transition kernel from $\bm{h}_{t}$ to $\bm{h}_{t-1}$ with a learnable parameter set $\bm{\theta}$, and $\bm{\mu}_{\bm{\gamma}}\left(\cdot \right)$ and $\bm{\Sigma}_{\bm{\gamma}}\left(\cdot \right)$ denote the mean and variance, respectively.

Based on variational inference (VI) framework, the evidence lower bound (ELBO) of $p\left(\bm{h}_{0} | \bm{h}_{0}^{r}, v_{s} \right)$ can be calculated as 
\begin{align}
\hspace{-0.2cm}\log p\left(\bm{h}_{0} | \bm{h}_{0}^{r}, v_{s} \right) \ge &  \nonumber\\
& \hspace{-2.5cm} E_{q}\left[\log p\left(\bm{h}_{T} | \bm{h}_{0}^{r}, v_{s}  \right) + \sum_{t=1}^{T} \log \frac{p_{\bm{\gamma}}\left(\bm{h}_{t-1} | \bm{h}_{t}, \bm{h}_{0}^{r}, v_{s}\right) }{q\left(\bm{h}_{t} | \bm{h}_{t-1}, \bm{h}_{0}^{r}, v_{s} \right)  } \right],
\setcounter{equation}{28}
\label{eq18}
\end{align}
where $E_{q}\left[\cdot \right]$ denotes the expectation for the joint distribution of latent variables with respect to $q\left(\bm{h}_{t} | \bm{h}_{t-1}, \bm{h}_{0}^{r}, v_{s} \right)$. Ignoring the constant term, the optimization function in terms of Eq. (29) is derived as
\begin{equation}
\min_{\bm{\gamma}}\sum_{t=1}^{T} D_{KL}\left[q\left(\bm{h}_{t} | \bm{h}_{t-1}, \bm{h}_{0}^{r}, v_{s} \right) || p_{\bm{\gamma}}\left(\bm{h}_{t-1} | \bm{h}_{t}, \bm{h}_{0}^{r}, v_{s}\right) \right],
\setcounter{equation}{30}
\end{equation}
where $D_{KL}\left(\cdot || \cdot \right)$ denotes the Kullback-Leibler (KL) divergence operation.

According to Bayes theorem and Markov property, $q\left(\bm{h}_{t} | \bm{h}_{t-1}, \bm{h}_{0}^{r}, v_{s} \right) $ can be calculated as 
\begin{equation}
q\left(\bm{h}_{t} | \bm{h}_{t-1}, \bm{h}_{0}^{r}, v_{s} \right) = \frac{q\left(    \bm{h}_{t} | \bm{h}_{t-1}, \bm{h}_{0}^{r}, v_{s} \right) \cdot q\left(\bm{h}_{t-1} | \bm{h}_{0},  \bm{h}_{0}^{r}, v_{s} \right) }{q\left(\bm{h}_{t} | \bm{h}_{0}, \bm{h}_{0}^{r}, v_{s} \right) },
\setcounter{equation}{31}
\end{equation}
where 
\begin{equation}
q\left(\bm{h}_{t} | \bm{h}_{t-1}, \bm{h}_{0}^{r}, v_{s} \right) \sim \mathcal{N}\left(\bm{h}_{t} | \bm{h}_{t-1} + \alpha_{t}\Delta_{h}, \kappa_{s}^{2}\alpha_{t}\mathbf{I}  \right),
\setcounter{equation}{32}
\end{equation}
\begin{equation}
q\left(\bm{h}_{t-1} | \bm{h}_{0}, \bm{h}_{0}^{r}, v_{s} \right) \sim \mathcal{N}\left(\bm{h}_{t-1} | \bm{h}_{0} + \eta_{t-1}\Delta_{h}, \kappa_{s}^{2}\eta_{t-1}\mathbf{I}  \right),
\setcounter{equation}{33}
\end{equation}
\begin{equation}
q\left(\bm{h}_{t} | \bm{h}_{0}, \bm{h}_{0}^{r}, v_{s} \right) \sim \mathcal{N}\left(\bm{h}_{t} | \bm{h}_{0} + \eta_{t}\Delta_{h}, \kappa_{s}^{2}\eta_{t}\mathbf{I}  \right).
\setcounter{equation}{34}
\end{equation}
Substituting Eqs. (32)-(34) into Eq. (31), the derivation yields
\begin{equation}
q\left(\bm{h}_{t-1} | \bm{h}_{0}, \bm{h}_{0}^{r}, v_{s} \right) \sim \mathcal{N}\left(\bm{h}_{t-1} | \frac{\eta_{t-1}}{\eta_{t} }\bm{h}_{t} + \frac{\alpha_{t} }{\eta_{t} }\bm{h}_{0}, \kappa_{s}^{2}\frac{\eta_{t-1} }{\eta_{t} }\alpha_{t}\mathbf{I}  \right).
\setcounter{equation}{35}
\end{equation}
So $\mu_{\bm{\gamma}}\left(\bm{h}_{t}, \bm{h}_{0}^{r}, v_{s}\right)$ can be expressed as 
\begin{equation}
\mu_{\bm{\gamma}}\left(\bm{h}_{t}, \bm{h}_{0}^{r}, v_{s}\right) = \frac{\eta_{t-1} }{\eta_{t} }\bm{h}_{t} + \frac{\alpha_{t}}{\eta_{t}}f_{\bm{\gamma}}\left(\bm{h}_{t}, \bm{h}_{0}^{r}, v_{s}, t  \right),
\setcounter{equation}{36}
\end{equation}
where $f_{\bm{\gamma}}\left(\cdot \right)$ denotes the model  built on DNN with parameter set $\bm{\gamma}$ that aims to predict $\bm{h}_{0}$.

So the optimization function shown in Eq. (30) can be equivalently expressed as training a neural network to fit the real channel distribution, i.e.,
\begin{equation}
\min_{\bm{\gamma}} \sum_{t=1}^{T}\frac{\alpha_{t} }{2\kappa_{s}^{2}\eta_{t}\eta_{t-1} }\left\|f_{\bm{\gamma}}\left(\bm{h}_{t}, \bm{h}_{0}^{r}, v_{s}, t\right) - \bm{h}_{0}  \right\|_{2}^{2}.
\setcounter{equation}{37}
\end{equation}

\begin{figure}[t]
\centering
\resizebox{0.49\textwidth}{0.18\textheight}{\includegraphics{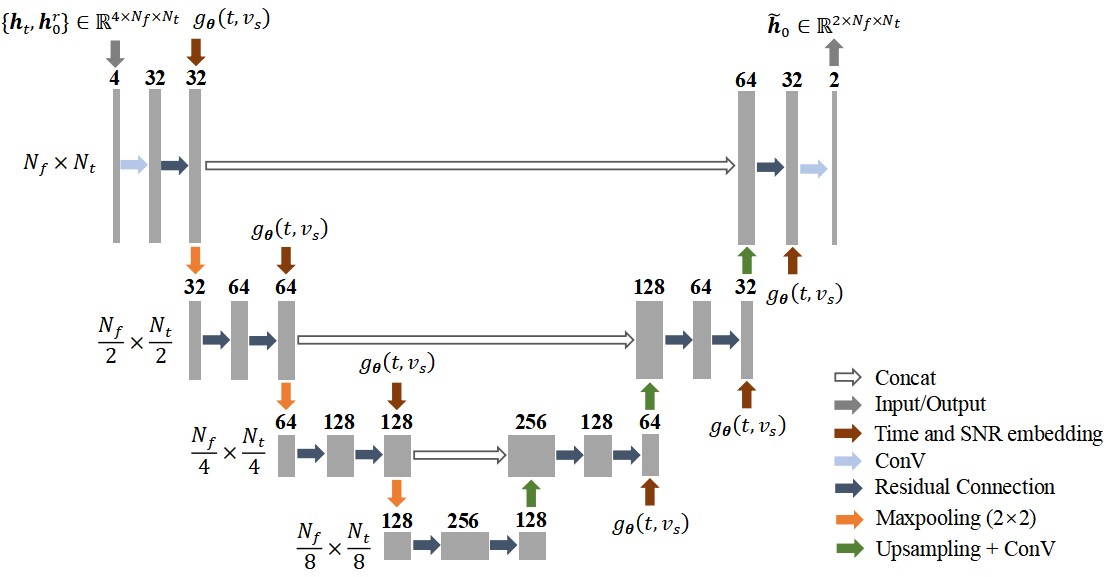}}
\captionsetup{font={footnotesize}}
\caption{The architecture of designed conditional U-Net.}
\label{fig4}
\end{figure}

\textit{3) Designed Conditional U-Net for Channel Prediction}  

As shown in Fig. 3, the conditional U-Net is designed to predict $\bm{h}_{0}$ along with the embedded time-step $t$ and SNR $ v_{s}$. 
Next, we introduce the gate fusion mechanism to adaptively adjust the weight of the time and SNR information.
The embedding of both the time-step and SNR value employs the positional encoding method proposed in Transformer framework [34]. As to the time-step, the embedding can be expressed as 
\begin{equation}
T_{emb}(t) = \left[\cos\left(t\cdot f_{0} \right), \sin\left(t\cdot f_{0}  \right), ...    \right]\in \mathbb{R}^{d_{T}}, 
\setcounter{equation}{38}
\end{equation}
where $f_{k} = 10000^{-\frac{2k}{d_{T}}}, k=0,1,...,\frac{d_{T}}{2}-1$, and $d_{T}$ denotes the embedded dimension of time information.

As to the SNR value, the embedding can be expressed as
\begin{equation}
S_{emb}(v_{s}) = \left[\cos\left(v_{s}^{norm}\cdot \gamma_{0} \right), \sin\left(v_{s}^{norm}\cdot \gamma_{0}  \right), ...    \right]\in \mathbb{R}^{d_{S}}, 
\setcounter{equation}{39}
\end{equation}
where $v_{s}^{norm}=\frac{\log_{10}^{\left(v_{s} + \epsilon\right)} - \log_{10}^{\epsilon} }{\log_{10}^{v_{s}^{max}} - \log_{10}^{\epsilon}}$ with $v_{s}^{max}$ denoting the maximum SNR value, $\gamma_{i} = 10000^{-\frac{2i}{d_{S}}}, i=0,1,...,\frac{d_{S}}{2}-1$, and $d_{S}$ denotes the embedded dimension of SNR information.

To adaptively fuse the time-step embedding and SNR embedding, the gating mechanism function is expressed as
\begin{equation}
g_{\bm{\theta}}\left(t, v_{s}\right) = \omega_{t} \odot T_{emb}(t) + \omega_{v_{s}} \odot S_{emb}(v_{s}), 
\setcounter{equation}{40}
\end{equation}
where $\omega_{t}$ and $\omega_{v_{s}}$ denote the weights trained along the model with parameter set $\bm{\theta}$.

Along the contracting path of the U-Net, the $\left\{\bm{h}_{t}, \bm{h}_{0}^{r}\right\}$ are stacked as the training data, and $g_{\bm{\theta}}\left(t, v_{s}\right)$ is added as the conditional information of embedding time and SNR information. Then, $3\times 3$ convolutional kernels along with the sigmoid linear unit (SiLu) are adopted for feature extraction and non-linear transformation. Meanwhile, residual connection is introduced to exploit the spatial correlation of the feature maps. Furthermore, the $2\times 2$ max-pooling is adopted to reduce the resolution of feature maps by half while the corresponding number of feature maps increases continuously. 
Finally, during the expansive path, the down-sampled feature maps are progressively up-sampled to match the resolution of their corresponding layers on the left side. Concatenating features from both the left and right paths facilitates the effective integration of low-level and high-level features.
For clarity, the training process and inference process of the proposed scheme for CSI refinement are summarized in \textbf{Algorithm 1} and \textbf{Algorithm 2}, respectively.

\begin{algorithm}[t]
\begin{small}
    \renewcommand{\algorithmicrequire}{\textbf{Input:}}
    \renewcommand\algorithmicensure {\textbf{Output:} }
    \captionsetup{font={small}}
    \caption{Training Process of the CDM-based Scheme}
        1: \textbf{Input:} Training dataset $\left\{\bm{h}_{0}, \bm{h}_{0}^{r}\right\}$, and epoch number $E^{k}$ \\
        2: \textbf{for} $e=1$ : $E^{k}$ \textbf{do} \\
        3: \hspace{0.2cm} Sample $t$ from the uniform distribution $t\sim U[1,T]$; \\ 
        4: \hspace{0.2cm} Obtain noisy channel data $\bm{h}_{t}$ by (22); \\
        5: \hspace{0.2cm} Fuse the time-step embedding and SNR embedding by (40); \\
        6: \hspace{0.2cm} Train the conditional U-Net in terms of (37); \\
        7: \textbf{end for} \\
        8: \textbf{Output}: Trained model $f_{\bm{\gamma}}\left(\cdot \right)$ with the parameter sets $\bm{\gamma}$ and $\bm{\theta}$
\end{small}
\end{algorithm}

\begin{algorithm}[t]
\begin{small}
    \renewcommand{\algorithmicrequire}{\textbf{Input:}}
    \renewcommand\algorithmicensure {\textbf{Output:} }
    \captionsetup{font={small}}
    \caption{Inference Process of the CDM-based Scheme}
        1: \textbf{Input:} $\bm{h}_{0}^{r}$, $T$, $v_{s}$, $f_{\bm{\gamma}}\left(\cdot \right)$ \\
        2: \textbf{for} $t=T$ : $1$ \textbf{do} \\
        3: \hspace{0.2cm} Obtain $\bm{h}_{t-1}$ by (35);\\ 
        4: \hspace{0.2cm} Calculate $\mu_{\bm{\gamma}}\left(\bm{h}_{t}, \bm{h}_{0}^{r}, v_{s}\right)$ by (36);  \\
        7: \textbf{end for} \\
        8: \textbf{Output}: Estimated channel $\hat{\bm{h}}_{0}$
\end{small}
\end{algorithm}

\subsection{Training Strategy} 

The proposed VQ-DSC-R system is trained by using a three-stage strategy. In the first stage, we train the SE and SD along with the SQC directly, where the multipath fading channel and noise are not considered. 
The loss function can be represented as 
\begin{equation}
\mathcal{L}\left(\mathbf{S}, \hat{\mathbf{S}}, \bm{\alpha}, \bm{\beta}, \mathbf{\bm{C}} \right) = d\left(\mathbf{S}, \hat{\mathbf{S}} \right).
\setcounter{equation}{41}
\end{equation}
Since the SQC update scheme (described in Section III-B) enables the gradient to be back-propagated, there is no need for specially designed item with respect to codebook loss. In this way, the optimal model parameters $\left\{\bm{\alpha}^{*}, \bm{\beta}^{*}, \mathbf{\bm{C}}^{*}\right\}$ can be obtained with considering the semantic feature extraction, quantization and reconstruction.

In the second stage, the CDM-based scheme is introduced to refine CSI for mitigate multipath fading channel inference. The initial estimated channel and conditional information are used as the inputs to train CDM, and then the actual channel distribution can be predicted.
So, the optimal model parameters $\bm{\gamma}^{*}$ of CDM can be obtained. 

In the third stage, the entire scheme is jointly retrained. Specifically, the parameters of SE, SQC and CDM are frozen, while the parameters of SD are updated with considering multipath fading channel and noise. The loss function in the first stage is adopted to renew the model parameters $\bm{\beta}$ of SD under $\left\{\bm{\alpha}^{*}, \bm{\gamma}^{*}, \mathbf{\bm{C}}^{*}\right\}$. 
Through the three-stage training strategy, it balances semantic-feature learning, quantization alignment, and channel adaptation, enabling the system to generalize effectively across diverse practical scenarios.

\section{Experiments and Results Discussion}

This section evaluates the performance of the proposed VQ-DSC-R system considering OFDM-based transmission over wireless channels.

\subsection{Experiments Setup} 

\textit{1) Datasets, Training Environments and Evaluation Metrics:} To verify the effectiveness of VQ-DSC-R system, the widely representative datasets are selected for training and testing. Specifically, the images from the OpenImage dataset [35] cropped to $256\times256$-pixel size constitute the training dataset. 
Since it covers a wide range of scene and object types, it can ensure the model's ability to generalize. 
The images from the DIV2K (DIVerse 2K) dataset [36] cropped to $1024\times1024$-pixel size constitute the testing dataset, including 100 high-resolution images. Note that the proposed model can be applied to image transmission with various resolutions. 
The proposed scheme is implemented in PyTorch framework and optimized via AdamW optimizer with a learning rate of $2\times 10^{-4}$ and weight decay of $10^{-4}$. The AdamW optimizer is worked with a batch size of 16 and a maximum iteration count of 100. All our experiments are performed on a Linux server with GTX 4090Ti GPU. Reconstruction fidelity is quantified to measure divergence between original and reconstructed images, including the PSNR for pixel-level accuracy [37] and MS-SSIM for perceptual quality assessment [38]. Bit compression ratio (BCR) is adopted to measure the required communication transmission costs. Specifically, BCR is defined as the ratio of the number of bits required for the transmission related to the original images, i.e., BCR=$\frac{B_{s} }{H\times W\times O\times 8 }$, where $B_{s}$ denotes transmission bits of the index sequence in terms of quantitative mapping. 

\textit{2) OFDM Symbol Structure and Channel Model:} The PRB of OFDM-based system encompasses a total of 1024 sub-carriers of each OFDM symbol and 14 OFDM symbols. The uniform grid-type pilot pattern is adopted, where the pilot intervals along the OFDM subcarriers and the OFDM symbols are set to be 9 and 5, respectively. To mitigate inter-symbol inference effectively, the CP length is set to be 256, ensuring it exceeds the delay spread of the multi-path channel. The 3GPP standard channel model is adopted to generate channel coefficients and added as multipath channel interference when training and testing our proposed model. Specifically, the EPA (extended pedestrian A) channel model [39] is designed for low-speed mobile scenarios, which can practically simulate the multipath fading of signals through scatterers such as buildings and trees in urban micro or indoor hotspots. \textbf{Table I} lists some other parameter conﬁgurations in the system, where ns denotes the delay in nanosecond unit.

\begin{table}[t]
\caption{Parameter Configurations}
\begin{tabular}{|l|p{3.5cm}|}
\hline
\textbf{Parameters} & \textbf{Values} \\
\hline
Training epoch & 100 \\
\hline
Training equipment & GTX 4090Ti GPU \\
\hline
Number of OFDM subcarriers & 1024 \\
\hline
Number of OFDM symbols & 14 \\
\hline
Pilot Pattern & Uniform grid-type \\
\hline
Pilot interval along the subcarriers & 9 \\
\hline
Pilot interval along the symbols & 5 \\
\hline
EPA Channel Delays (ns) & [0, 30, 70, 90, 110, 190, 410] \\ 
\hline
EPA Channel Powers (dB) & [0, -1.0, -2.0, -3.0, -8.0, -17.2, -20.8] \\
\hline
\end{tabular}
\end{table}

\textit{3) Comparison Schemes:} To validate the advantages of VQ-DSC-R, different scenarios and schemes are implemented for comparison with other baseline schemes. 
\begin{itemize}
\item[]
\hspace{-0.5cm} $\bullet$
\hspace{0.1cm}\textit{SQC Update Schemes:} 
The popular STE scheme is proposed to deal with issue of the non-differentiable nearest-neighbor assignment, which copies the gradients through a non-differentiable VQ function. In this case, the codebook loss and commitment loss are integrated with the reconstruction loss to optimize the encoder, codebook and decoder. The NSVQ scheme simulates the VQ distortion by adding noise to the latent vector to mimic the original quantization error. In this way, the quantized vector is a differentiable function of the latent vector and the selected codeword. These two typical schemes represent the main ideas for handling quantitative mapping, whose impacts on image transmission are thoroughly studied in our VQ-DSC-R system. 

\hspace{-0.5cm} $\bullet$
\hspace{0.1cm}\textit{Channel Estimation Schemes:} 
The least square (LS) scheme is widely used to obtain CSI at pilot position, based on which the linear interpolation is used to obtain the CSI at data position [40]. The deep residual channel estimation network (ReEsNet) is proposed in [41] for channel estimation, which outperforms many other deep learning-based estimation schemes.
The performance of these two schemes are compared with that of our proposed CDM-based scheme in terms of the normalized mean squared error (NMSE) and bit error rate (BER).
Meanwhile, the impacts of CSI on image transmission of VQ-DSC-R system are studied. The performance of image transmission under the perfect CSI is considered as the upper limit for comparson. The 4 quadrature amplitude modulation (4-QAM) is considered as the modulation scheme without loss of generality. 

\begin{figure}
\centering
\subfigure[]{
  \label{fig:subfig:a}
   \includegraphics[width = 0.38\textwidth]{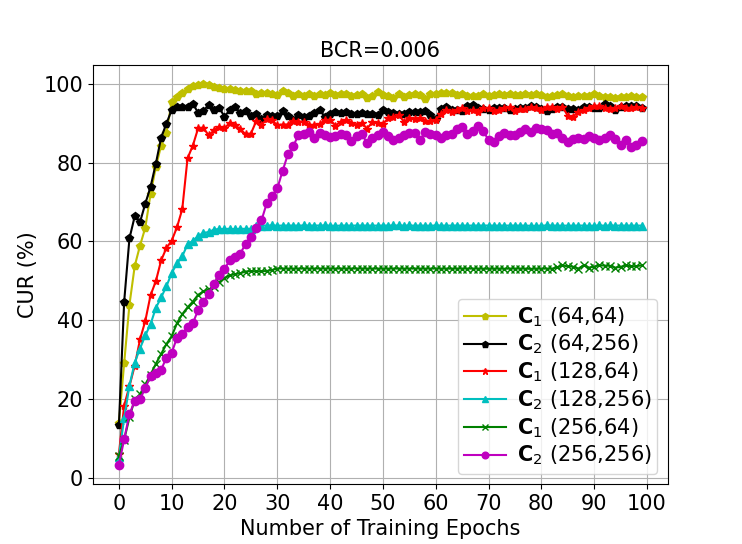}}
\hspace{1in}
\subfigure []{
 \label{fig:subfig:b}
 \includegraphics[width = 0.38\textwidth]{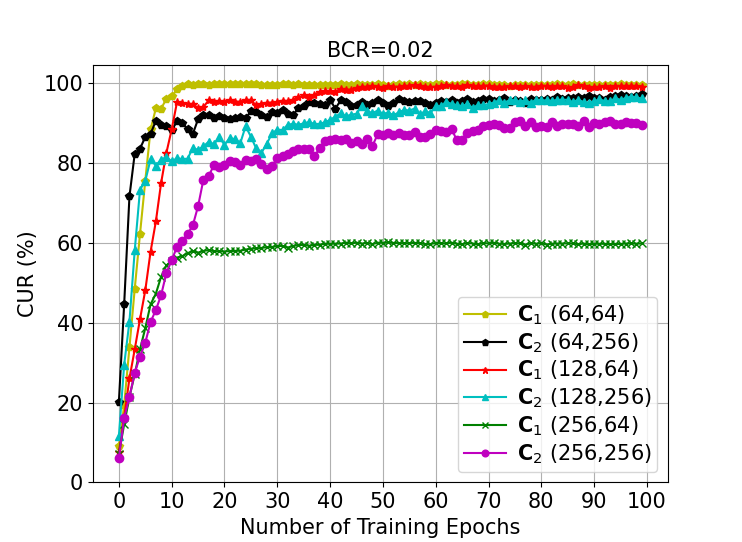}}
\captionsetup{font={footnotesize}}
\caption{CUR performance curves for the ANDVQ scheme versus the number of training epoches. (a) BCR=0.006; (b) BCR=0.02.} 
\label{fig:subfig}
\end{figure}

\hspace{-0.5cm} $\bullet$
\hspace{0.1cm}\textit{Conventional  SSCC Scheme and DJSCC Scheme:}
As a conventional baseline for comparison, the popular joint photographic experts group (JPEG) and low density parity check (LDPC) are used for source encoding and channel encoding, respectively.
As the conventional SSCC scheme fails to recover images under imperfect channel estimation, it is considered with perfect CSI. Additionally, the ADJSCC model [31] is adopted for comparison in terms of SNR adaptation. To enable compatibility with our proposed VQ-DSC-R system, we redesign ADJSCC by fusing its SNR-adaptive module with our VQ-DSC-R system for joint semantic extraction and quantization, which is referred to as VQ-ADJSCC. 
\end{itemize}

\begin{table}[t]
\centering
\captionsetup{font={footnotesize}}
\caption{PSNR and MS-SSIM Performance Comparison of SQC Update Schemes With Different Codebook Sizes and BCRs.}
\label{tab:codebook_performance}
\begin{tabular}{>{\small}l|>{\small}c|>{\small}c>{\small}c|>{\small}c>{\small}c}
\toprule
\textbf{Method} & \textbf{K} & \multicolumn{2}{>{\small}c|}{\textbf{BCR=0.006}} & \multicolumn{2}{>{\small}c}{\textbf{BCR=0.02}} \\
& & \textbf{PSNR} & \textbf{MS-SSIM} & \textbf{PSNR} & \textbf{MS-SSIM} \\
\midrule
ANDVQ & 64 &  24.62 & 0.872 &  27.88 & 0.944 \\
      & 128 & \textbf{25.66} & \textbf{0.888} & \textbf{29.22} & \textbf{0.959} \\
      & 256 & 25.33 & 0.887 &  28.57 & 0.950 \\
\midrule
STE   & 64 &  20.71 & 0.726 & 21.07 & 0.782 \\
      & 128 & 20.75 & 0.726 & 22.73 & 0.850 \\
      & 256 & 20.32 & 0.691 & 22.41 & 0.832 \\
\midrule
NSVQ  & 64 &  21.25 & 0.736 & 22.78 & 0.834 \\
      & 128 & 22.59 & 0.840 & 23.57 & 0.856 \\
      & 256 & 22.02 & 0.800 & 23.24 & 0.832 \\
\bottomrule
\end{tabular}
\end{table}

\subsection{Performance of SQC Update Schemes} 
In this part, numerical simulations are provided to investigate the performance of the proposed SQC update scheme, i.e., the ANDVQ scheme, as well as compare with the STE scheme and NSVQ scheme. 
To evaluate the performance of trained SQC, 
the multipath fading channel and additive noise are ignored.
For the two-level quantitative mapping design presented in Fig. 2(a), each codebook's size $K$ is manually set and each codebook's semantic codeword dimension $N$ is configured to match the channel dimension of the feature maps. 
So the BCR can be calculated accordingly.
The BCR=0.006 and BCR=0.02 are chosen for simulation. Note that these two BCR values are approximations.
As to the BCR=0.006, the parameters of SE and SD are set as $n=4, O_{1}=16, O_{2}=64, O_{3}=256, O_{4}=256, O_{5}=64, O_{6}=32$ and $O_{7}=16$.
As to the BCR=0.02, $n=2$ is set without changing other parameters of SE and SD. Other hyperparameters of ANDVQ scheme are set as $K_{c}=5$, $\gamma_{0}=0.9$ and $\epsilon=10^{-5}$.

\begin{figure}
\centering
\subfigure[]{
  \label{fig:subfig:a}
   \includegraphics[width = 0.38\textwidth]{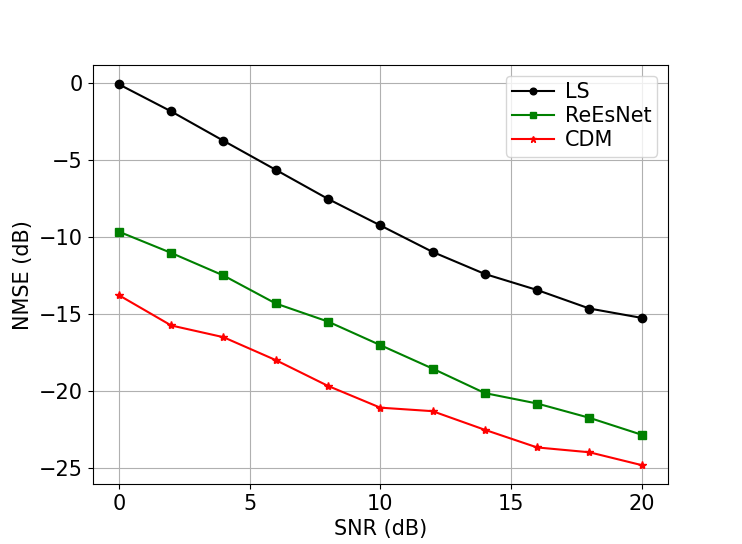}}
\hspace{1in}
\subfigure []{
 \label{fig:subfig:b}
 \includegraphics[width = 0.38\textwidth]{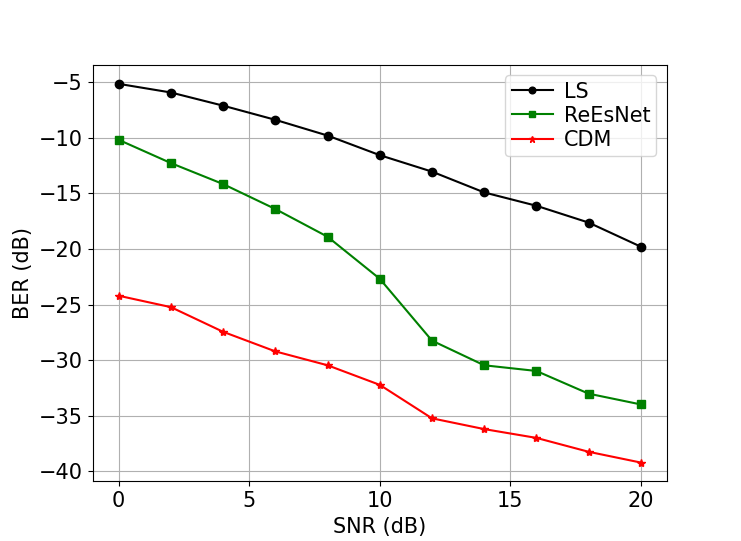}}
\captionsetup{font={footnotesize}}
\caption{NMSE and BER performance curves for different schemes versus different SNR conditions. (a) NMSE Performance; (b) BER Performance.} 
\label{fig:subfig}
\end{figure}

\begin{figure*}[t] %

\subfigure[]
{
\begin{minipage}[c]{5.5cm}
\centering
\includegraphics[width=1.0\linewidth]{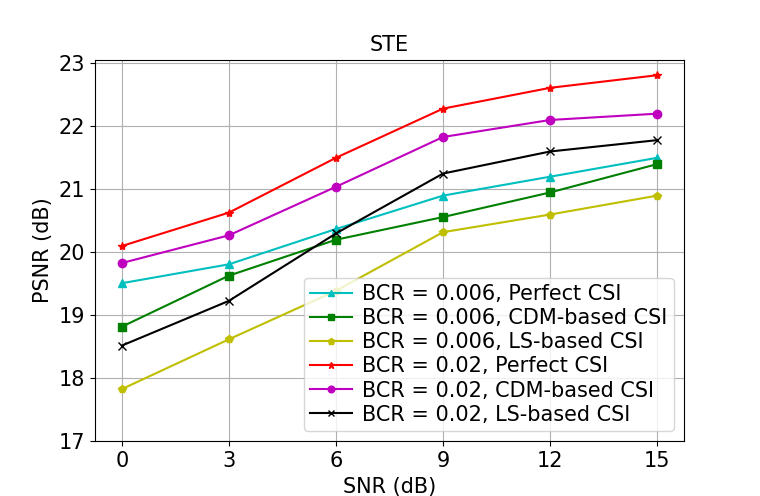}
\end{minipage}
}
\subfigure[]
{
\begin{minipage}[c]{5.5cm}
\centering
\includegraphics[width=1.0\linewidth]{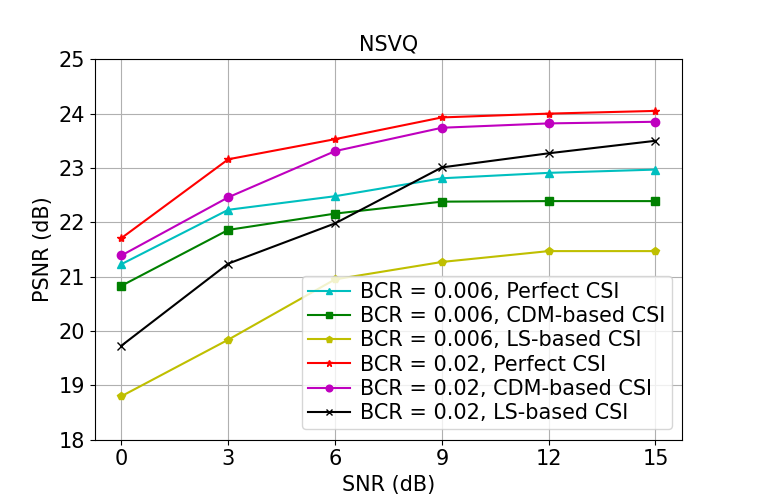}
\end{minipage}
}
\subfigure[]
{
\begin{minipage}[c]{5.5cm}
\centering
\includegraphics[width=1.0\linewidth]{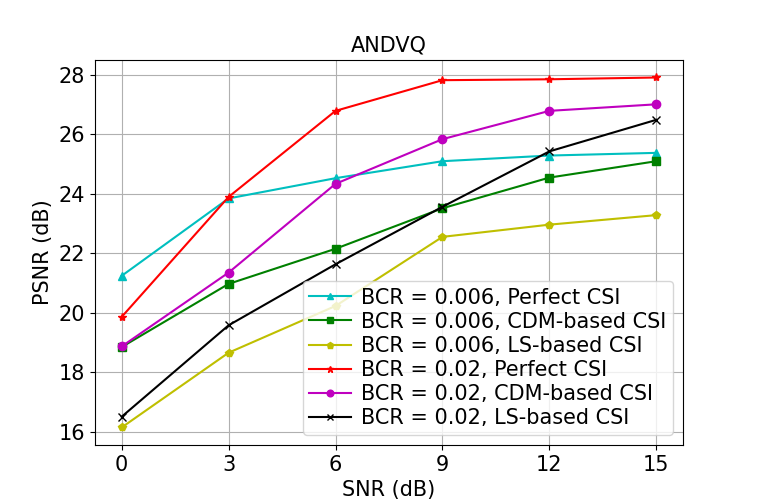}
\end{minipage}
}
\subfigure[]
{
\begin{minipage}[c]{5.5cm}
\centering
\includegraphics[width=1.0\linewidth]{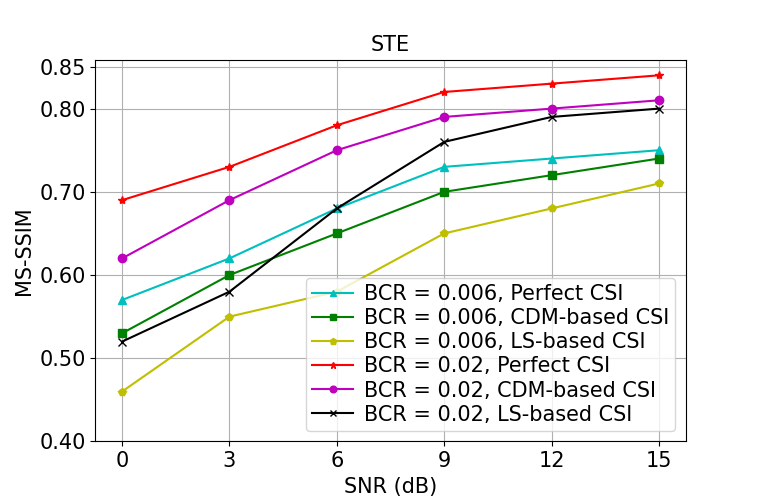}
\end{minipage}
}
\subfigure[]
{
\begin{minipage}[c]{5.5cm}
\centering
\includegraphics[width=1.0\linewidth]{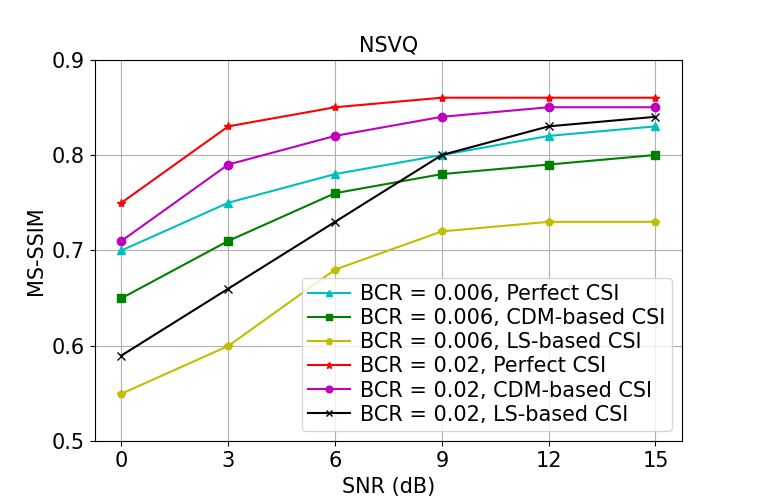}
\end{minipage}
}
\subfigure[]
{
\begin{minipage}[c]{5.5cm}
\centering
\includegraphics[width=1.0\linewidth]{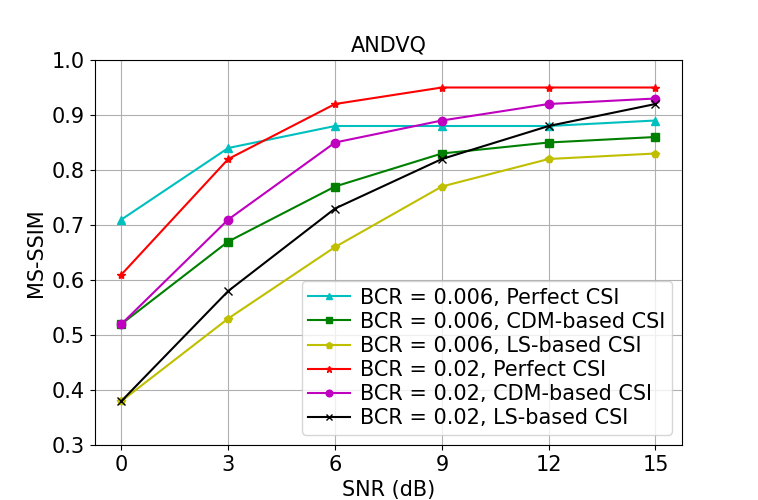}
\end{minipage}
}
\caption{PSNR and MS-SSIM performance of SQC update schemes under different channel estimation schemes and BCRs. (a) and (d) present the performance of STE scheme; (b) and (e) present the performance of NSVQ scheme; (c) and (f) present the performance of ANDVQ scheme.}
\label{figs}
\end{figure*}

Fig. 4 presents the codebook utilization ratio (CUR) of our proposed SQC update scheme versus the number of training epochs. 
The CUR, which is defined as the proportion of the actual number of codebook entries used in the quantization process to the total codebook capacity, reflects the representational capability of the codebook to a certain extent.
As to different codebook sizes $K=\{64, 128, 256\}$ and BCR=$\{0.006, 0.02\}$, the CURs increase as the number of training epochs, and all of them have achieved very high CURs.  
It's worth noting that the CUR with $K=64$ is approaching a saturated state, which may result in codebook capacity overfitting. 
Additionally, Table II presents the PSNR and MS-SSIM performance achieved by different SQC update schemes with different codebook sizes and BCRs. It can be observed that: 1) Compared with the STE scheme and NSVQ scheme, the ANDVQ scheme obtains the best performance in terms of PSNR and MS-SSIM. 
As to BCR=0.006, the PSNR and MS-SSIM obtain 25.66 and 0.888, respectively. As to BCR=0.02, the PSNR and MS-SSIM are increased to 29.22 and 0.959, respectively. 
2) As the number of codebook size increases, the performance of codebook capacity doesn't always improve consistently. 
Meanwhile, the codebook with $K=128$ obtains the best performance compared to those with other codebook sizes. So it means that the codebook size and CUR should be considered simultaneously for evaluating codebook capacity.

\begin{figure}
\centering
\subfigure[]{
  \label{fig:subfig:a}
   \includegraphics[width = 0.38\textwidth]{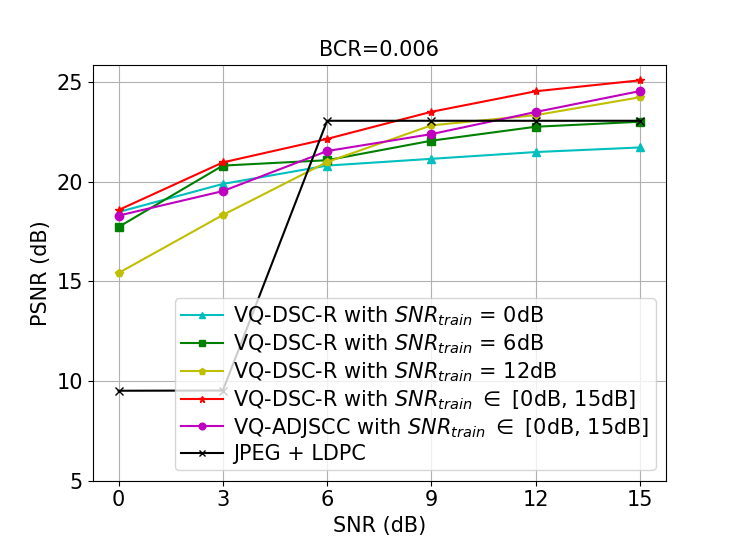}}
\hspace{1in}
\subfigure []{
 \label{fig:subfig:b}
 \includegraphics[width = 0.38\textwidth]{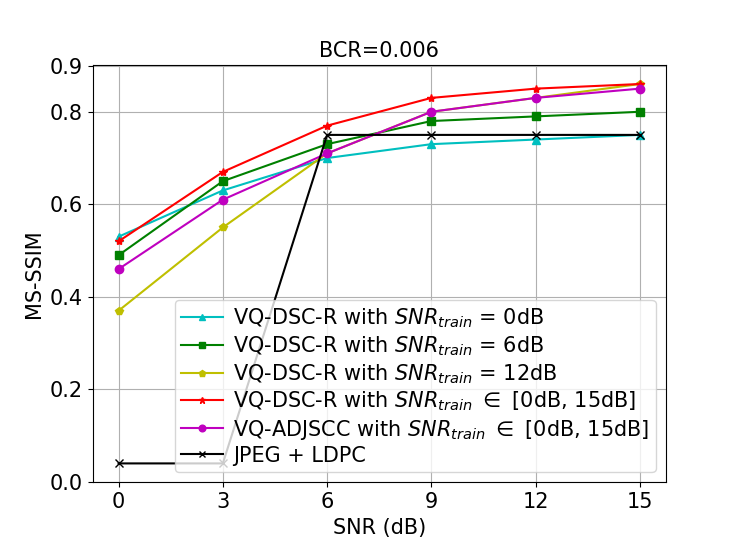}}
\subfigure []{
 \label{fig:subfig:b}
 \includegraphics[width = 0.38\textwidth]{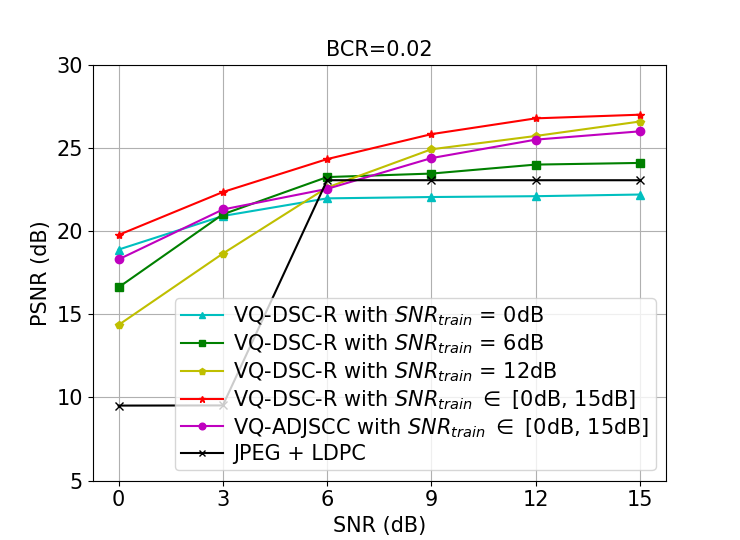}}
\subfigure []{
 \label{fig:subfig:b}
 \includegraphics[width = 0.38\textwidth]{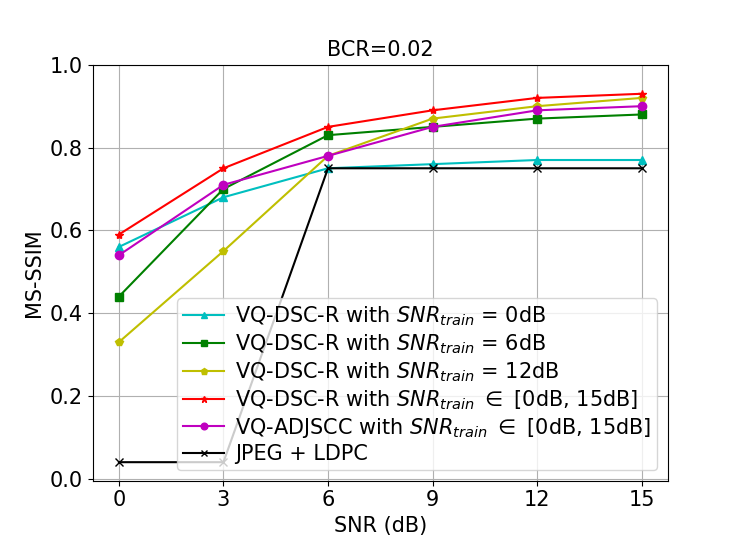}}
\captionsetup{font={footnotesize}}
\caption{Channel adaptive performance of VQ-DSC-R under different SNR and BCRs. (a) and (b) present the performance with BCR=0.006; (c) and (d) present the performance with BCR=0.02;} 
\label{fig:subfig}
\end{figure}


\subsection{Performance of Channel Estimation Schemes} 

Fig. 5 presents the NMSE and BER performance of the proposed CDM-based scheme under various SNRs compared with those of LS-based scheme and ReEsNet-based scheme. 
The hyperparameters of CDM-based scheme are set as $\kappa_{0}=1$, $\rho_{0}=0.5$, $E^{k}=100$, $v_{s}^{max}=20dB$, $d_{T}=d_{S}=10^{4}$. For the forward diffusion process, the step of adding noise is set as 20. For the reverse sampling process, the step of inference is set as 5. 
Fig. 5(a) shows that the CDM-based scheme can significantly decrease NMSE compared with the LS-based scheme under various SNRs. Meanwhile, it outperforms the ReEsNet-based scheme due to its powerful ability of noise reduction. 
Additionally, Fig. 5(b) provides the BER performances under various SNRs, which clearly indicate the impacts of inaccurate CSI on index transmission.

By setting codebook size $K=128$, Fig. 6 provides the PSNR and MS-SSIM performance of SQC update schemes under different channel estimation schemes and BCRs. It can be observed that: 1) As the SNR increases, the PSNR and MS-SSIM performance increase, and the  ANDVQ scheme obtains much better performance compared with those of STE scheme and NSVQ scheme. 2) Compared with the model trained with LS-based CSI, the performance with CDM-based CSI is better, and the performance with perfect CSI is the best. The results clearly demonstrate the impact of channel estimation error on end-to-end training and testing, which should be considered in practical scenarios.

\subsection{Performance of SNR adaption} 
The VQ-DSC-R model integrating  \textit{SNR ModNet} is trained on signals with SNR uniformly sampled from range [0dB, 15dB].  
For benchmarking purposes, we employ three standalone VQ-DSC-R models lacking \textit{SNR ModNet}, and each model is trained specifically at one fixed SNR level (0dB, 6dB, or 12dB).
Additionally, the traditional SSCC model (i.e., JPEG and LDPC) and popular ADJSCC model are chosen for comparison. 
Fig. 7 shows the PSNR and MS-SSIM comparison for VQ-DSC-R, VQ-ADJSCC, and traditional scheme with different BCRs. Specifically, the BCR of traditional scheme is 0.1, and BCRs of other DL-based schemes are 0.02 and 0.006. 
It can be observed that: 
1) VQ-DSC-R outperforms traditional methods in PSNR and MS-SSIM, particularly at medium-low SNRs. Meanwhile, it exhibits a smooth performance degradation and avoids the \textit{cliff effect}.
2) VQ-DSC-R surpasses VQ-ADJSCC in performance by leveraging the proposed \textit{SNR ModNet}.
3) Higher $\mathrm{SNR_{test}}$ improves VQ-DSC-R's performance monotonically, and the VQ-DSC-R with \textit{SNR ModNet} maintains significant advantages over the \textit{SNR ModNet}-free baseline regardless of deviations between $\mathrm{SNR_{train}}$ and $\mathrm{SNR_{test}}$.

Fig. 8 presents some visualized examples of comparison between the reconstructed images of VQ-DSC-R schemes and original images under multipath fading channel model and different SNRs. Two different types of images are selected.
From left to right are the reconstructed images of the original image with the increasing of SNRs, and from the top to bottom are the reconstructed images of the original image with the increasing the BCRs. Compared with other schemes, the proposed VQ-DSC-R scheme not only improves the PSNR and MS-SSIM performance but also makes certain progress in visual perception.

\begin{figure*}[htbp]
\centering
\resizebox{0.95\textwidth}{0.50\textheight}{\includegraphics{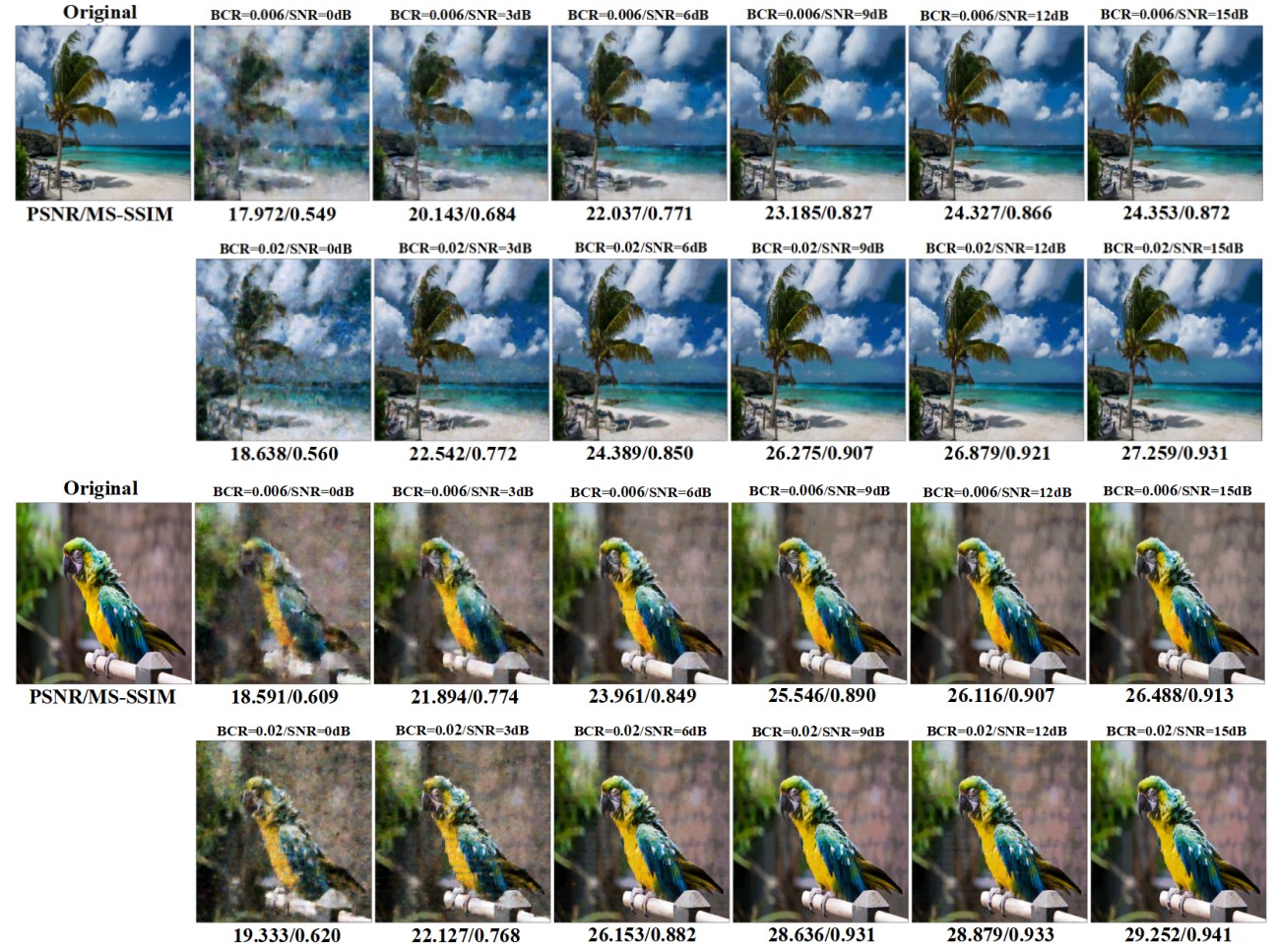}}
\caption{Two examples of comparison between the reconstructed images of VQ-DSC-R scheme and original images under EPA channel model and different SNR conditions. From left to right are the reconstructed images of the original image with increasing of SNR, and from the top to bottom are the reconstructed images of the original image with increasing of BCR.}
\label{fig1}
\end{figure*}

\section{Conclusion}

This paper introduces VQ-DSC-R, a robust digital semantic communication scheme that bridges semantic and digital communication via leveraging VQ mechanism for efficient image transmission over OFDM systems in wireless fading channels. 
Within the DJSCC framework, a Swin Transformer-based backbone extracts multi-scale semantic features, which are projected by VQ modules into a trainable SQC for efficient index-based transmission. 
The proposed ANDVQ scheme for SQC update dynamically adjusts quantization mapping by using K-nearest neighbor noise-variance statistics, which is integrated with EMA to ensure stable training. 
Furthermore, leveraging a CDM-based CSI refinement and an attention-guided channel adaptation module, the system adaptively optimizes index transmission over multipath fading channel and noisy environments. 
A three-stage training strategy jointly optimizes feature learning, quantization, and channel adaptation, ensuring generalization across scenarios. 
Extensive experiments demonstrate that VQ-DSC-R achieves superior semantic performance with high compression ratios under practical OFDM transmission conditions.

\end{document}